%% file: main.tex
\documentclass[letterpaper,twocolumn,10pt]{article}
\usepackage{usenix2024_SOUPS}

\usepackage{tikz}
\usepackage{amsmath}

\usepackage{multirow}
\usepackage[flushleft]{threeparttable}
\usepackage{csquotes}

\newcommand\peri[1]{\textcolor{magenta}{#1}}
\newcommand\Jeff[1]{\textcolor{blue}{#1}}
\newcommand\KIvy[1]{\textcolor{red}{#1}}
\newcommand\chris[1]{\textcolor{green}{#1}}

\usepackage{filecontents}

\begin{document}
%-------------------------------------------------------------------------------

%don't want date printed
\date{}

\title{\Large \bf Privacy or Transparency? Negotiated Smartphone Access as a Signifier of Trust in Romantic Relationships}

\def\plainauthor{Periwinkle Doerfler and Kieron Ivy Turk and Chris Geeng and Damon McCoy and Jeffrey Ackerman and Molly Dragiewicz}

\author{
{\rm Periwinkle Doerfler}\\
New York University
\and
{\rm Kieron Ivy Turk}\\
University of Cambridge\\
New York University
\and
{\rm Chris Geeng}\\
New York University
\and
{\rm Damon McCoy}\\
New York University
\and
{\rm Jeffrey Ackerman}\\
Griffith University
\and
{\rm Molly Dragiewicz}\\
Griffith University
}

\maketitle
\thecopyright % Ivy: I modified the usenix...sty file to remove SOUPS copyright but keep author copyright.

%-------------------------------------------------------------------------------
\begin{abstract}
%-------------------------------------------------------------------------------
In this work, we analyze two large-scale surveys to examine how individuals think about sharing smartphone access with romantic partners as a function of trust in relationships.
We find that the majority of couples have access to each others' devices, but may have explicit or implicit boundaries on how this access is to be used. 
Investigating these boundaries and related social norms, we find that there is little consensus about the level of smartphone access (i.e., transparency), or lack thereof (i.e., privacy) that is desirable in romantic contexts. However, there is broad agreement that the level of access should be mutual and consensual. Most individuals understand \textit{trust} to be the basis of their decisions about transparency and privacy. 
Furthermore, we find individuals have crossed these boundaries, violating their partners' privacy and betraying their trust. We examine how, when, why, and by whom these betrayals occur. 
We consider the ramifications of these boundary violations in the case of intimate partner violence.
Finally, we provide recommendations for design changes to enable technological enforcement of boundaries currently enforced by trust, bringing access control in line with users' sharing preferences.
\end{abstract}

\input{intro.tex}
\input{background.tex}

\input{methodology.tex}
\input{sharing.tex}
%\input{privacy.tex}
%\input{transparency.tex}
\input{trust}

\input{betrayal.tex}
\input{discussion.tex}
\input{conclusion}

\section*{Acknowledgements}
This study was funded in part by NSF grants 1916126 and 2016061.

%-------------------------------------------------------------------------------
%\section*{Acknowledgments}
%-------------------------------------------------------------------------------

%Grant funding, etc.
%Kevin Roundy at Symantec

%-------------------------------------------------------------------------------
\bibliographystyle{plain}
\bibliography{survey-bib}

\input{appendix}
\end{document}

%% file: intro.tex
\section{Introduction}
% %\chris{condensing}
% Smart phones are increasingly central fixtures in our modern digital lives. For many, 
% %parts of the developing world coming into the internet era "mobile first." \cite{}
% %For many, 
% phones are the primary medium used to consume news and entertainment, connect with others through texts, calls, and social media, conduct business via email and online banking, navigate to destinations, and shop. %\cite{} 
% Smartphones create great convenience by aggregating our online presence on one device, allowing users to log in to each service once and have ready access to it via its application.

% The structure of mobile operating systems, however, is such that sharing access to one's phone is essentially a binary: someone has access or they do not. 
% If a person has access to the device for innocuous matters such as looking up directions or playing a game, they can also read texts, access social media, investigate search histories, and more.
% Applications for credentialed services, for example email or social media, generally only require a login once, remaining permanently authenticated thereafter.

85\% of US adults own a smartphone, and an increasing number of people depend on them for Internet access~\cite{mobilefactsheet}. %As tools to communicate, learn, and find news, smartphones create great convenience by aggregating our online presence on one device, allowing users to log in to each service once and have ready access to it via its application. 
Phone access design is mostly a binary: someone has access or they do not. If a person has access to the device for innocuous matters such as looking up directions or playing a game, they can also read texts, access social media, investigate search histories, and more.
%Although many couples share smartphone access, sharing access is designed to be mostly binary: someone has access or they do not. If a person has access to the device for innocuous matters such as looking up directions or playing a game, they can also read texts, access social media, investigate search histories, and more. 
There are many circumstances where sharing access is necessary, convenient, or desirable, but few tools exist that allow people who trust each other to set their ideal sharing, security, and privacy settings on smartphones~\cite {wu2022sok}. This raises the tension between privacy and transparency in trusting relationships. Does one trust their partner so thoroughly as to grant them unfettered device access (i.e., provide transparency), or does their partner trust oneself so completely as to forego access to one's device altogether (i.e., prefer privacy)?

%Hence, handing someone an unlocked smartphone automatically places a good deal of trust with them, and there are many circumstances where this is necessary, convenient, or desirable. 
Researchers have found that families and couples commonly share devices and credentials~\cite{holt2021personal,sunny_sharing,caring_sharing,Park2018,lin2020did}. %Romantic couples share devices and accounts for “relationship maintenance, household maintenance, trust, and convenience”~\cite{Park2018}, as well as due to occupational and couple norms~\cite{lin2021s}. 
%Indeed, many people share their devices freely with those they trust. \cite{sunny_sharing, caring_sharing} 
%In the context of romantic relationships --- ideally grounded in trust --- a tension between  privacy and transparency emerges. 
%Does one trust their partner so thoroughly as to grant them unfettered access (i.e., transparency), or does one trust their partner so completely as to forego access altogether (i.e., privacy)? (check this last phase - I don't understand is this supposed to say *not* trust?)
%Computer science research has examined the ways people think about sharing their devices \cite{sunny_sharing,caring_sharing} and accounts. \cite{Park} 
Social psychological research has examined couples' expectations about using phones to communicate with each other~\cite{phone_conflict_autonomy, tech_ipv_contextual}, as well as with opposite-sex acquaintances~\cite{intercouple_agreement, phone_conflict_autonomy}. Little research, however, has investigated couples' expectations about using \textit{each others' devices.} 
Researchers %have also examined trust and digital privacy in couples, but this work has largely focused on 
have also investigated the relationship between trust and `snooping' behaviors amongst couples~\cite{trust_mitigates_snooping, distrust_leads_snooping,snooping_capacity_matters, surveillance_uncertainty_attachmentstyle, FBaddiction_leads_surveillance}, as well as prevalence of intimate partner monitoring~\cite{intercouple_agreement,addl_snoop_ipv_stats}.
%There is a significant body of computer and social science literature on intimate partner monitoring, measuring its prevalence  \cite{intercouple_agreement,addl_snoop_ipv_stats} and examining factors ranging from attachment style to tech literacy. \cite{snooping_capacity_matters, surveillance_uncertainty_attachmentstyle, FBaddiction_leads_surveillance} %Much of this work, however, uses poorly defined constructs and assumes that ambiguous sharing behaviors are abusive without exploring the context in which they occur, the meaning of behaviors, or their impact on those involved. \cite{Brown2018} 
Much of this work, however, assumes that ambiguous sharing behaviors are negative without exploring the context in which they occur, the meaning of behaviors, or their impact on those involved~\cite{Brown2018}. 
%While some family studies research assumes that any degree of privacy (or desire for it) is problematic~\cite{read_dms_ok}, the meaning, implications, and outcomes of privacy and transparency are entirely context-dependent. Behaviors used to build intimacy in some relationships may be coercive or controlling in others~\cite{tech_ipv_contextual,Dragiewicz2021}. For example, young people in peer groups rife with infidelity may find transparency a healthy method of trust building~\cite{infidelity_perception_monitoring}.

To investigate couples' device-sharing behaviors within relationships, we present findings from two surveys: one survey conducted by the authors (1) measuring participants' access to intimate partners' devices and (2) assessing perceptions of normative and acceptable use of that access, 
and another survey conducted by Norton, (3) measuring non-consensual access to partners' devices.  Our main findings are:

%e conducted a large-scale online survey to measure the prevalence of  individuals granting partners access to their phones and assess attitudes about %Jeff's comment - perhaps "the normative and acceptable use..." .
%normative and acceptable use of that access. 

%We find significant evidence that relationships evolve towards access parity even when partners may disagree about what would be optimal.  We find some differences in behaviors and expectations across age.
%Additionally, we provide the first granular analysis of survey data from Symantec to examine how often people abuse their access to their partners' devices, and draw on both datasets to understand why.
%We examine the extent to which access behaviors and expectations differ along demographic lines, finding few gender differences, but some age differences. 
\begin{enumerate}
    \item There is little consensus on what constitutes normative device-sharing behavior; every behavior we measured had strong proponents and staunch critics, almost always motivated by trust.
    \item Most people agree that access and usage behaviors should be negotiated, mutual, and consensual.
    \item The primary justification participants have for non-consensual device access is concerned about infidelity, though we note that there is a blurry line between ``justifiable'' or benign non-consensual access and abuse \cite{IPS_narratives}.
\end{enumerate}

Finally, we offer a set of recommendations for design changes in mobile operating systems and apps to allow granular access control, removing the stricture of binary access and making negotiated boundaries 
%(e.g. `you can have access to my phone, but don't read my texts') 
technologically enforceable.  
%We encourage future work to assess the impact of such changes on romantic relationships should they be adopted, as well as more detailed research examining the norms surrounding sharing behaviors. %This might include contextualized examination of the relationship between benign and abusive access to partners' devices, an analysis of the relationship between perceptions of sharing norms and personal sharing preferences, the impact of domestic violence on sharing and consent, and the influence of digital media and social media on privacy and infidelity norms.

%thoughts about trust and access fall on a spectrum

%explicit v implicit understandings 

%dichotomous access leads to boundaries etc
%understanding of where boundaries

%fact that boundaries are interpersonal not enforced means they are breached

%how when why that happens

%ta da maybe we should change that

%% file: background.tex
\section{Related Work}

\subsection{Sharing in Intimate Relationships}

Researchers have explored how families and other close relations think about and practice device and credential sharing~\cite{holt2021personal,sunny_sharing}. While this practice is common between people who trust each other~\cite{sunny_sharing}, amongst couples it can lead to unintentional privacy concerns~\cite{caring_sharing}.
%as well as issues with revoking account access when a relationship ends~\cite{Park2018}. %Through in-depth interview and memory-journal work, Matthews et al. demonstrate that people frequently share devices with those they live with and trust, often without significant consideration of the implications of that access~\cite{sunny_sharing, hol}. %In similar work, Jacobs et al. enumerate 
%\Jeff{just pointing out that you're switching the verb tense in this sentence (present) versus the prior sentence (past) - best to keep these consistent} %the unintentional privacy concerns that emerge from convenience-driven sharing in couples~\cite{caring_sharing}.
%Park et al. studied account sharing in different stages of romantic relationships~\cite{Park2018}, with as trust grows partners share more of passwords and devices, but at relationships end, revoking access becomes tedious. 
Reasons for account sharing in romantic couples include ``relationship maintenance,
trust, and convenience''~\cite{Park2018}, as well as occupational and couple norms~\cite{lin2021s}. Couples living together in a smart home also navigate sharing smart device access~\cite{geeng2019s,zeng2019understanding}. Despite device and credential sharing being common, Wu et al. find that few tools allow intimate partners to maintain both their ideal social and security behaviors in this context ~\cite{wu2022sok}.

%Holt et al. explore family thoughts on sharing and security concerns around post-mortem data and pre-planning before death~\cite{holt2021personal}

While much computer security literature on password sharing assumes it should be discouraged or prevented %or some novel scheme like biometric fuzzing. 
~\cite{password_sharing_bad, password_fuzzing}, recent research finds cases where digital security and privacy are managed cooperatively~\cite{hayes2019cooperative}. 
Kaye studies a small group of individuals to understand password sharing broadly across apps, finding that the behavior is both common and well-considered~\cite{kaye_2011}.
Singh et al. study sharing of banking passwords, finding that sharing is common and convenient in couples and sometimes completely unavoidable for individuals with disabilities~\cite{banking_passwords}. This sharing within couples applies to other accounts as well, whether through passwords, joint accounts, or leaving devices unlocked~\cite{caring_sharing}. %Though there are some studies --- from market research as well as academia --- which measure password sharing for specific services such as Netflix~\cite{netflix}, there is a dearth of large-scale data about device and credential sharing in general. 

\subsection{Privacy and Trust in Intimate Contexts}
Device and credential-sharing research begins to touch on the issue of privacy in intimate contexts. In psychology, \textit{Communication-Privacy Management} theory explores the creation of shared privacy boundaries in families, where multiple people coordinate managing private information~\cite{petronio2010communication}. Specifically in romantic relationships, the \textit{Connection-Autonomy Dialectic} explains the tension between intimacy and freedom. Duran et al. apply this framework to couples' negotiated expectations of digital communication, %finding that couples more easily overcome disagreements about failing to answer phone calls or frequent texting than about communication with opposite-sex acquaintances. 
finding that some people provide partners access to their devices to reduce conflict~\cite{phone_conflict_autonomy}.

In a study interviewing married heterosexual couples in the UK, Hesper et al. find that husbands and wives do not always agree about norms for appropriate online communication with opposite-sex acquaintances. In couples where snooping occurs, they find that both partners generally do so equally~\cite{intercouple_agreement}. %Sambasivan et al. examine the technological and performative strategies South Asian women use to protect their privacy on their mobile devices in a cultural environment that discourages women's privacy in their communications~\cite{Sambasivan}.
Laborde et al.'s study of Latino youth finds that the perception of rampant infidelity means that access to partners' devices and low levels of digital privacy within couples can be a beneficial way to build trust in relationships~\cite{infidelity_perception_monitoring}.

\subsection{Monitoring and Snooping}

There is extensive research examining digital behaviors in romantic relationships, though few studies have investigated whether people in relationships consider monitoring behavior expected or problematic. %Little research, however, has investigated whether these behaviors are considered normative or deviant. 
%For example, Hester et al.'s work mentioned above inquires about couples' understanding of normative behavior and about snooping, but not about the intersection between the two. The assumption is that a disagreement about whether it is acceptable to watch online pornography or complain about your wife to a friend may lead that wife to snoop. 
%
%Research in social and computer science investigates the prevalence of online snooping and monitoring of partners and factors contributing these behaviors.
Researchers have found several factors that contribute to partner snooping, including distrust in a relationship~\cite{distrust_leads_snooping}, confidence in the technological capacity to gain the information they desire without getting caught~\cite{snooping_capacity_matters}, and relationship attachment style, though relationship uncertainty was not an influencing factor~\cite{surveillance_uncertainty_attachmentstyle}.
%In regards to factors contributing to snooping between partners, Fox et al. identify a correlation between individuals' attachment style and propensity to snoop: relationship uncertainty was not an influencing factor~\cite{surveillance_uncertainty_attachmentstyle}.
%Arikewuyo et al. find that distrust in a relationship is a predictor of snooping~\cite{distrust_leads_snooping}. Tokunaga et al. find that individuals are much more likely to snoop if they are confident in their technological capacity to gain the information they desire without getting caught in the process~\cite{snooping_capacity_matters}. 

Partner monitoring research also looks at partners monitoring their partners' social media accounts \textit{without accessing their devices}~\cite{SNS_conflict_notinfidelity,SNSaddiction,FBaddiction_leads_surveillance,FB_jealousy_selfesteem,FB_maintenance_jealousy}. %investigating relationship factors that predict this behavior and the behavior as a predictor of relationship problems.
% Arikewuyo et al. find a significant correlation between levels of social media usage and partner monitoring, which leads to relationship conflict. Notably, they do not find that social media usage is correlated with infidelity~\cite{SNS_conflict_notinfidelity}.
% However, Abbasi et al. find that levels of usage they classify as addiction are correlated with social media-related infidelity~\cite{SNSaddiction}.
% Elphinston et al. find that social media addiction is a predictor of partner monitoring, and in turn, relationship jealousy and conflict~\cite{FBaddiction_leads_surveillance}. 
% Utz et al. find that monitoring is predictive of jealousy for people with low self-esteem~\cite{FB_jealousy_selfesteem}, and Anderegg et al. find that both monitoring and transparency are predictive of jealousy~\cite{FB_maintenance_jealousy}.
%This research has been conducted by researchers who study relationships rather than social media, however, and 
Many of these studies view social media usage as a cause of relationship problems; however, some research is contradictory. Clayton finds that the tenure of a relationship has no impact on whether Twitter use becomes a point of conflict~\cite{twitter_breakup}, while Smith finds that Facebook use is only correlated with negative outcomes in relatively new relationships~\cite{FB_usage_generation}.
%Furthermore, some studies find underlying relationship issues to lead to social media monitoring, while others find social media monitoring leads to relationship issues. %While there is a correlation between the two, the direction of causality remains unclear.
%Given the direction of causality remains unclear, longitudinal research could help understand whether social media use is independently detrimental to relationships, or merely showcases underlying relationship issues.%a new platform by which underlying issues manifest.

Security researchers have studied ``extreme cases'' of relationship monitoring using spyware. Chatterjee et al. identify hundreds of apps available on app stores with stalking capabilities, including some actively advertising partner monitoring as a use case~\cite{rahul}. Bellini et al. discuss the fora where users of this software congregate, and how they justify this behavior~\cite{IPS_narratives}. Freed et al. examine difficulties in remediating technologically facilitated abuse including spyware, as well as victims' concern about abusers' reactions to lost access~\cite{nyc_ipv}. Wei et al. study the kinds of ``anti-security'' advice TikTok users recommend to surveil partners~\cite{wei2022anti}.

\subsection{Abuse of Technology in Coercive Control}
There is a sizable body of literature on the role of technology in coercive control\footnote{Coercive control is a pattern of abusive behaviour that includes physical and non-physical abuse, facilitated by structural gender inequality. See Stark~\cite{Stark_2013} for more information.} \cite{Dragiewicz2021, eSafetykids2020, ACCAN,Stark_2013,Woodlock_2013,Woodlock_2015,WESNET2ndnatlsurv,Douglasetal2019,Woodlock_McKenzie_Western_Harris_2020}, %Much of this work involves surveys with professionals who work with survivors of domestic violence \cite{Woodlock_2013,Woodlock_2015,WESNET2ndnatlsurv} and interviews with survivors about the role technology plays in abuse~\cite{ACCAN, eSafetykids2020}. This work grew out of increasingly frequent anecdotal findings on technology from studies on other aspects of domestic violence~\cite{Harris_Woodlock_2019, Douglasetal2019}. 
demonstrating digital media and mobile phone enabled abuse.
Freed et al. %examine technological abuse from a cybersecurity perspective, framing 
framed abusers as \textit{authenticated but adversarial} device users; this type of adversary %undermines traditional threat models and 
emphasizes the need for designers to contemplate interpersonal threats~\cite{stalker_paradise}.
Transparency and privacy on mobile phones present significant concerns for abuse survivors. Woodlock finds that 56\% of women's abusive partners use mobile technology to check their location, 47\% check their text messages without permission, and 17\% demand electronic passwords~\cite{Woodlock_2017}. Douglas et al. find that abusers demand to inspect women's phones~\cite{Douglasetal2019,ACCAN}. Dragiewicz et al. report that abusers also pressure children to provide login information for their mothers' phones~\cite{ACCAN,eSafetykids2020,DragiewiczMumPW}.
While this work is critically important, it captures only technology-facilitated abuse occurring in relationships already identified as abusive by an affected person. 
More research is needed to understand the continuum from normative technology sharing to serious abuse; our study seeks to fill this gap related to normative sharing side of the continuum. %We find that for all types of access behaviors measured, there are some people who find them desirable. Accordingly, more detailed work is needed to understand whether the presence of specific behaviors can be regarded as indicative of abuse or control, or whether the meaning of transparency and privacy behaviors is completely context-driven. 

%\subsection{Consumer Attitudes About Digital Privacy}
%This is for work that is not specifically about intimate contexts or about sharing, just the further point that people *do not* think about their privacy that much, are not really willing to do much to protect it, etc. 

%% file: methodology.tex
\section{Methodology}
%Our survey contained closed and open-ended questions. 
%We recruited through a combination of Facebook ads and Prolific. 
%Here, we provide an overview of the survey design, participant recruitment, and analyses performed. 
%See Section \ref{sec:datasets} for an overview of sample demographics. 

\subsection{Data Collection}
Our survey was fielded from November 2020 to March 2021 in two phases. First, we recruited participants using targeted ads on Facebook. Second, we used Prolific to recruit.

\subsubsection{Facebook Recruitment}
We chose to recruit through targeted Facebook ads because we were conscious of the fact that users registered with survey administration platforms like Mechanical Turk and Prolific tend to be of higher socio-economic status (SES) and more familiar with technology than the population as a whole~\cite{pew_mtruk}. We understood that both of these factors could influence individuals' propensity to share their device with a partner, e.g., due to financial constraints or the need for technical assistance. Furthermore, because these platforms require dedicated registration, there is an additional bias imposed by the stagnant nature of the user base. While ``Facebook users'' is also a relatively fixed population, 70\% of US adults use Facebook~\cite{fbusers}. 
%Further, the primary business model of the platform is targeted ads, so we felt this would be an effective method to recruit participants who are not registered with microwork sites.

Facebook advertising is most effective when targeted~\cite{fbads}.
%so while we targeted all adults using smartphones, we ran many small campaigns instead of one broad one. 
We ran campaigns dedicated to recruiting men, but none dedicated only to women, as women are over-represented among likely survey respondents and Facebook users~\cite{genderbias_survey,fbusers,response_bias}. Similarly, we did not target high-income individuals, and sometimes explicitly excluded college graduates. We also targeted the LGBTQ+ community 
\footnote{Ads targeted by sexual orientation are not allowed. A common proxy is to target fans of pages that are popular in the LGBTQ+ community such as Lady Gaga or RuPaul's Drag Race.} to reach LGBTQ+ users. 
%successfully recruiting a sample which was 12\% queer.
%We further reduced the reach of some campaigns by targeting only users that were interested in surveys, SurveyMonkey, and other related topics. This targeting, combined with the nature of Facebook's algorithms, certainly facilitates a self-selection bias. However, we argue that seeking individuals who are interested in surveys should not be worse in this regard than a dedicated survey-taking site requiring independent registration. Our ads did not generate a random sample of Facebook users, instead compensating for known selection biases and demographic skew by over-targeting demographic groups less likely to respond to surveys.

%\Jeff{targeting individuals in ways that would produce a sample that was purposely not representative of the population of mobile phone users - "bias" has a specific meaning in statistics and this wouldn't fit that meaning in this context}

We had intended to conduct the entirety of our participant recruitment via Facebook advertising but after several user reports on our ads (which we have since verified \textit{were not} in violation of Facebook's ad policies), our ad account was suspended by Facebook's monitoring systems. The first two times we asked to have it reinstated, it was. The third time, it was not, and we needed to identify another avenue of recruitment. 327 participants came from Facebook. \footnote{Meta has confirmed that recruiting survey participants is not in violation of its advertising policies, and has since created a dedicated channel for academics to leverage ads for this purpose allowing for direct support.}
%While all sampling has bias, we feel confident that the sample of participants recruited from Facebook is at least as representative of a sample recruited from platforms like mturk or prolific. 

\input{table_demo}

\subsubsection{Prolific Recruitment}
After losing access to Facebook advertising, we recruited the second half of our sample through Prolific. 
Prolific respondents were recruited in two batches: one seeking only men, 
%as women were over-represented in our participants from Facebook, 
and one seeking only individuals whose household income was less than \$60,000 and who did not have a Bachelor's degree, 
both in an effort to balance the existing sample. See Table \ref{tab:demo} for details on the demographics of our sample across the two recruitment strategies.  

\subsection{Survey Design}
Our survey included four main sections: 1) technical questions such as whether and how participants' phones lock, 2) questions about participants' partners, their technology, and the participants' access to it, 3) questions about perceptions of certain technological behaviors in a relationship context, 
%both with respect to whether they were good and whether they were common, 
and 4) an abuse screener and demographic questions. 
Single respondents did not answer questions in section 2, or the abuse screener in section 4. 

\subsubsection{Compensation}
%\Jeff{I made some alterations here} 
We collected no personally identifiable information (PII) from participants recruited from Facebook. Instead, upon survey completion respondents were given a unique code and directed to our website for a chance to win a \$50 Amazon gift card. Gift cards were distributed to one in twenty-five participants, and winners were notified immediately upon entry. %Winners were given a code to paste into their Amazon account, with no personal information collected. 
This scheme was designed to allow for the distribution of compensation without collecting any PII, as would have been necessary for direct cash payments like PayPal. %We opted for Amazon gift cards as these were the nearest cash replacement we could think of which could be distributed anonymously. 

It was not possible to purchase gift cards in \$2 increments, so we conducted a drawing. The amortized compensation was \$2 per respondent. The structure of compensation for Facebook respondents was made explicitly clear in both the advertisements and at the beginning of the survey. 
It is unfortunately not possible to avoid collecting Prolific IDs, %. These are collected and used in conjunction with attention tests and timers to 
which are used to validate attention tests and timers and approve payments. 
Once the entries were validated and payments approved, we did not retain these identifiers. Prolific respondents were paid a flat rate of \$2 to complete our survey.

\subsubsection{Question Design}
The survey included questions around participant's access and usage of their partner's phone, and vice versa.
%In section 1, questions were yes/no or multiple choice, e.g., \textit{``Does your partner know your phone PIN?''} Section 2 included the same questions as section 1 but in reverse, e.g., \textit{``Do you know your partner's phone PIN?''} as well as questions measuring the participant's usage of their partner's phone.
% %Questions were also added to measure how often the participant engaged in certain acts on their partner's phone. These were provided as a series of behaviors with the question \textit{``How often would you do the following?''} 
%Answer choices included \textit{Sometimes, if my partner asks me to} and \textit{Sometimes, if my partner is not in the room}, in addition to \textit{Never, Rarely, and Often.}
Section 3 presented participants with a series of behaviors ranging from \textit{``Sharing a bank account with a partner''} to \textit{``Checking a partner's location history on their phone''}.
First, participants were asked whether the behavior is \textit{common} in the context of romantic relationships, on a scale of \textit{very uncommon} to \textit{very common.} Then, participants were asked to make a moral judgment about the same behaviors: were they \textit{good/healthy, normal, problematic, or bad/toxic?}

%Some of the demographic questions were asked in a manner that is non-standard; instead of asking participants to provide a household income, we asked them to choose from one of five descriptive statements about their financial situation, e.g., \textit{``I live paycheck to paycheck.''} We chose this because the same income can imply vastly different lifestyles depending on one's circumstances and geography.
%%\Jeff{Peri - check this word "qualitative" - in most social science literature that pertains to surveys, this term refers to narratives - while in alternative literature it can mean a set of answers that are categorical in nature rather than numeric/interval - I don't know if this really matters for a conference paper - but I think it would be safer to use a word other than "qualitative" unless you're speaking about answers where the respondent types in a written narrative}  
%about household finances instead of asking for household income, as the same dollar figure can mean wildly different lifestyles for different people 
% \Jeff{alterations here} 
%who live in different locations.
%We also provided four descriptive statements to assess familiarity with technology, e.g. \textit{``I am quite familiar with technology. I use advanced settings and complicated software.''} 
%\Jeff{same here}

%\Jeff{made alterations here}
To evaluate whether a participant was in a romantic relationship, we first asked about marital status. 
For respondents who indicated they were never married, were divorced, or were widowed, we provided statements such as \textit{``I am currently dating or going out with someone''} to determine whether they were in any meaningful form of a romantic relationship regardless of the terminology the participant might use to describe that relationship, given %. We did this instead of simply asking participants if they were in a relationship, because 
%\Jeff{made alterations here} 
individuals define relationships differently and that terminology is both gendered and generational~\cite{10.2307/26347959}.
%Participants who indicated that any of the statements applied 
%\Jeff{made alterations here} 
%as well as those who indicated that they were married or engaged were asked questions pertaining to a partner. 
The complete text of the survey questions and answers is available in the supplemental materials.

\subsection{Ethical Concerns}
%\chris{moving this around so easier to find}
%\Jeff{I added some language here to be more specific} 
Our survey included questions about behaviors that could be viewed as negative within a relationship context, as well as a set of questions about behaviours that many would consider serious partner violence. Due to the sensitivity of these questions and the way they differed from the rest of the survey, participants were warned about their nature at the start of this section and advised they may skip it if they believed it would make them too uncomfortable. 
%an abuse screener (which participants were able to skip without being asked any questions) which could have made people uncomfortable. 
There are ethical concerns with using microwork platforms such as Prolific, including that they may be exploitative~\cite{mturk_ethics}.
In our Prolific recruitment, we are cognizant of this and provided a survey reward that equated to an hourly wage that exceeded the US minimum wage. Our institutions' IRBs approved this research.

\subsection{Analysis}
This paper includes an analysis of our data and data collected by Norton.
Here, we discuss our sample and how analyses were conducted. Table~\ref{tab:demo} provides an overview.

\subsubsection{Inclusion for Purposes of Calculations}
Because several sections of the survey were administered only to participants with a partner 
%the survey consisted of several sections with differing eligibilities 
%\Jeff{this term seems awkward to me how about, "because many sections of the survey were administered only to individuals for whom the questions in these contingent sections were relevant (based upon previous answers) the number of participants who answered any given question varies.}
and all questions were optional, the number of respondents who answered any given question varies. 
Respondents who failed attention check questions~\footnote{e.g.,``What planet do you live on?''~\cite{attn_check}}, 
finished the survey unreasonably quickly, gave the same answer choice for all questions, or answered fewer than 75\% of the questions %they were administered 
\footnote{Participants were not penalized for declining to answer abuse-related questions.}
were removed from our analyses. 

The remaining participants are included in the analyses related to the questions they answered, meaning that the reported number of respondents for each question differs slightly. Thus, when we report \textit{N} for a question, ``\textit{N} respondents'' means ``\textit{N} respondents who were shown and answered this question''. 
Some questions included ``I don't know'' %or ``I don't know what this is'' 
as an answer choice. Participants who chose this have been removed from calculations on these items.% though on occasion we report what fraction chose this answer if it is informative. 

\subsubsection{Demographics}
Demographic questions, except age, were optional. 
%As a result, there are respondents who dutifully answered most of our questions, but for whom we do not have demographic data. 
When making comparisons between groups, statistics are calculated including only respondents who provided demographic information and an answer to the question at hand.
When reporting aggregate statistics, respondents without demographic information are included; this is why, in some tables, the N for all demographic categories is not equal to N for all participants. %As aforementioned, N will be reported with all statistics.

While we are respectful of the fluidity of gender---the response options for gender identity included ``Non-Binary'' and ``Other''---the nature of the statistical analysis is such that it is not possible to draw meaningful conclusions for groups with few respondents. As such, when making gendered comparisons, we include only a comparison of those who identified as ``Female'' and ``Male''. 

We collected age as a continuous variable, but we have performed most analyses with age in brackets. 
%In the interest of having sufficient numbers of respondents in each age bracket, we chose to use fewer brackets than a typical bracketing in decades, instead using four brackets. 
We have split age approximately along canonical generational boundaries: ``Gen Z'', those born after 1995, or in 2021, those under age 26 (though we excluded those under 18), ``Millenials'', currently aged 26 - 40, ``Gen X'', currently aged 41-59, and "Baby Boomers," those 60 years or older. The maximum age in our sample was 82, and the minimum was 18. 
%We feel that these brackets are reasonable with respect to the subject matter of the study: the oldest member of Gen Z would have been 11 when the iPhone was invented and many will not remember a time without them; Millenials may or may not remember a time before Internet prominence, where Gen X will have been some ways into adulthood as the Internet approached ubiquity.

%We asked participants what the highest level of education they had achieved was, ranging from "no high school diploma" to "graduate degree." 
%We also asked participants about their financial circumstances. 

%we provided qualitative statements describing financial situations, as the same annual income can be meaningfully different in different parts of the country, and every household has unique constraints, resources, and expenses outside of strict income.
%We wanted to understand how social class might impact respondents behavior and attitudes with respect to interpersonal digital privacy. The answer choices were:
%\begin{itemize}
%    \item {\textit{"I struggle to pay my bills or put food on the table. Money is a real problem for me."}}
%    \item {\textit{"I live paycheck to paycheck."}}
%    \item {\textit{"I don't worry about my bills, but there's not much extra. I  have enough savings for an emergency."}}
%    \item {\textit{"I have everything I need and sizeable savings. I could travel or help my children pay for college."}}
%    \item {\textit{"I have everything I need and no debt. I would be able to stop working and still live comfortably."}}
%\end{itemize}

In our analysis, there are many places where we make comparisons among participants with different relationship statuses. While we understand that every relationship is different and marriage is not a unilateral indication of a couple's commitment level, we have defined ``serious'' relationships as those participants who were married, engaged, in a civil union, domestic partnership, or common-law marriage~\footnote{Often this comparison is made between married and unmarried couples, however, to include more forms of committed couples we include other categories: 14\% of respondents in "serious" relationships were LGBTQ+.}.
%Prior work intending to evaluate the most well-established of couples has often considered only married heterosexual couples, \cite{intercouple_agreement} but we have included other categories in the interest of inclusivity; 14\% of respondents deemed to be in "serious" relationships identified as members of the LGBTQ+ community. 
We did not ask participants about the duration of their relationship. 
Any individuals deemed to have a partner, but not in the above categories were considered to be in a ``less serious'' relationship.  
Additionally, we asked individuals whether their relationship was long-distance or online-only, and if so, did not include them in our questions about their partners' technology as the lack of physical proximity would preclude access to each others' phones.

\subsubsection{Norton Data}
In addition to our survey data, we also analyze Norton's complementary survey data on partner monitoring, which Norton commissioned for market research in 2019 to inform mobile anti-virus software design, to emerge risks from stalkerware and spyware in IPV contexts. We perform the first extensive analysis of these data.
%Norton undertook this research to inform product design for mobile anti-virus software, with an eye towards emerging risks from stalkerware and spyware in IPV contexts but had not yet investigated potential broader findings in their data.
The survey was administered by Harris Poll\footnote{\url{https://theharrispoll.com/}}, included 2,050 respondents---1,200 female and 850 male respondents---and was approximately representative of the US population concerning age, education, and household income. 
%and reflective of Norton's likely consumer for the product they were designing.

Norton asked respondents exclusively about \textit{non-consensual} behaviors, including monitoring (``online stalking'' or ``Facebook stalking''), snooping on their partner's device, logging into their accounts on other devices, and tracking through location services or dedicated applications.
They asked respondents why they had done those things, as well as about their attitudes towards online stalking. %Aggregated response statistics are available in supplemental materials.

\subsection{Statistical Tests}

We use logistic regressions to check for correlations between responses and demographic factors. % using logistic regressions to identify the impact of different demographic factors on user behaviour. 
To ensure that the assumptions of logistic regressions are met---primarily, a large sample size for each demographic, and having no excessive multicollinearities where independent variables are correlated with each other
---we remove Baby Boomers and non-binary people's responses from the models and remove identified multicollinear factors. We use chi-squared tests to identify significant factors in these models and for additional tests. %For some additional tests not covered by the regressions, we continue to use chi-squared tests.

\subsection{Limitations}
While the Norton dataset is representative of the US population, our Prolific and Facebook data are not. %survey data recruited from Prolific and Facebook is not representative of the US population. 
Demographic questions (except age) were optional. Our survey design does not look at whether the relationship is monogamous, polyamorous, or somewhere in between, and does not ask about relationship duration. Long-distance partners may also engage in phone and account access when co-located. 
%We also did not ask about changes in sharing after break-ups. 
While our open-ended items allow for qualitative answers, we are not able to ask follow-up or clarifying questions. 

%% file: table_demo.tex
\begin{table*}[]
\centering
\begin{tabular}{l|l|lll|llll|lll}
         &     & \multicolumn{3}{l|}{Gender \& Orientation} & \multicolumn{4}{l|}{Age}    & \multicolumn{3}{l}{Relationship Status} \\
         & N   & Female        & Male        & LGBTQ+       & 18-25 & 26-40 & 41-59 & 60+ & Single       & Serious      & Other      \\ \hline
Total    & 531 & 274           & 225         & 63           & 196   & 174   & 121   & 40  & 154          & 193          & 184        \\ \hline
Facebook & 327 & 210           & 96          & 33           & 107   & 98    & 90    & 32  & 96           & 120          & 111        \\
Prolific & 204 & 64            & 129         & 30           & 89    & 76    & 31    & 8   & 58           & 73           & 73        
\end{tabular}
\caption{Breakdown of demographic factors of sample across recruitments}
~\label{tab:demo}
\end{table*}

%% file: sharing.tex
\section{Sharing and Access}
%Handing someone an unlocked phone requires trust, but giving them ongoing independent access requires more; they cannot only read messages, but with some types of access may also be able to open banking apps, authorize purchases, or even install malicious apps.
In this section, we examine the access individuals have to partners' phones, whether that access is knowledge-based or biometric, and whether the access is reciprocal within the relationship. 
In Section \ref{sec:trust} we examine the role of interpersonal trust in the choice of whether to grant access and in Section \ref{sec:betrayal} we examine how access can be and is abused.

\input{5_1}

\subsection{Partner Access}
After establishing that respondents' phones locked, we asked whether their partner had \textit{independent access} through each authentication scheme in use, i.e., 
%if the respondent used a PIN, we asked if their partner knew it, 
if the respondent used facial recognition, we asked whether their partner could unlock their phone using \textit{their own face}. 
%We asked respondents the same questions about their access to their partner's phone.
If an individual's phone did not lock, we assumed their partner had access to it. 
If someone indicated they did not know whether their partner's phone locked, we assumed they did not have access to it. 

\subsubsection{Access and Demographic Factors}
\label{sub_sub:access_demo}
Of participants with partners (N = 378), 13.8\% indicate that their phone does not lock, granting implicit access, and another 50.8\% indicate that their partner has explicit, independent access. Asked about partners' phones, 14.1\% of respondents (N = 341) said their partner's phone doesn't lock, and a further 51\% indicate that they have access.
Of 189 respondents whose phones lock and partners have access, 95.2\% have given their partners their SYK, either a PIN or swipe pattern. Significantly fewer respondents, only 27.3\%, offer their partner biometric access, including 4.8\% of respondents who report that their partner has \textit{only} biometric access. The figures were comparable for respondents' access to their partners' phone. We discuss demographic correlations with implicit and explicit phone sharing in Appendix~\ref{app:access}.
%See Table \ref{tab:access_t} for details.
%We find that biometric access is generally offered \textit{in addition} to knowledge-based access, with nearly a quarter of respondents granting both. There is not a significant difference in the rates at which men and women grant biometric access ($\chi^2(1, 324) = .66, p=.416$), nor between ``serious'' and ``less serious'' relationships ($\chi^2(1, 324) = .03, p=.862$).

%\input{table_access_type}

\subsubsection{Mutual Access}
%Nearly identical portions of respondents indicated that their partner had access to their device (64.6\%, N = 378) and that they had access to their partner's device (65.1\%, N = 341), however, this is aggregated. 
We also investigated whether access is mutual within relationships. 
For the 314 couples about whom we have complete information, we examine access in a pairwise capacity to check for symmetry.
%and there were virtually no demographic differences in the rates at which people granted their partner access, it is still prudent to 
%In the remainder of this section, 
%
Including explicit and implicit access, 55.4\% of respondents indicated that they and their partner both have access, while 23.5\% of respondents indicate that neither they nor their partner has access. These proportions are significantly higher than would be expected according to a binomial test comparing to an equal distribution (p<.0001), indicating that states of symmetrical access are the norm. Table \ref{tab:mutual_access} provides a breakdown of the four access situations. 

\input{table_mutual_access}

There is a significant difference between men and women in the distribution amongst the four possible states of access,
%for male and female respondents, 
with women being significantly more likely to be in a state of mutual access, and
men significantly more likely to have mutual non-access ($\chi^2(3, 341)=9.93, p=.019$).
However, amongst heterosexual couples, there is not a significant difference in the distributions for men and women ($\chi^2(3, 288)=5.62, p=.132$).

There is no significant difference between heterosexual and homosexual couples ($\chi^2(3, 341)=1.13, p=.770$), however, there is a significant difference between gay and lesbian couples, with 82\% of respondents in lesbian couples reporting mutual access, compared to 41\% of respondents in gay couples ($\chi^2(3, 44)=9.73, p=.021$).
While there are tendencies towards mutual access and mutual non-access, most demographic groups are not significantly more likely to be in a relationship with asymmetrical access than any other. The exception to this is that women are significantly more likely to have symmetric access than men. (Over generations, ($\chi^2(3, 341)=3.43, p=.330$, men/women $\chi^2(3, 332)=9.93, p=.019$ queer/not: $\chi^2(1, 341)=.19, p=.663$).
Of the 61 respondents in a heterosexual relationship with asymmetrical access, in 27 the woman had access and in 34 the man did. This difference is not significant ($\chi^2(1, 61)=1.61, p=.205$).
%As noted in \ref{sub_sub:access_demo}, future work emphasizing same sex couples could elucidate the true role of gender preferences with respect to these behaviors, without the influencing factor of gendered parity in relationships.
This further supports findings in \ref{subsub:lock_demo} and \ref{sub_sub:access_demo} in suggesting that behavior within a relationship tends to have parity as opposed to imbalance. 
%Further work focusing on same-sex couples could be useful in understanding the extent to which there may be genuine gendered preferences regarding access, mitigated by parity.

%% file: 5_1.tex
\subsection{Locking and Authentication}~\label{subsub:lock_demo}
To understand whether respondents' partners had access to their phone (and vice versa) we first asked respondents whether their phones locked. We assume that if one's phone does not lock, anyone with whom they regularly share physical space will have implicit access to it. For individuals whose phones locked, we asked what form of authentication they used. 
%\subsubsection{Phone Locking and Demographic Considerations}
Most participants (83.7\% , N = 552) said that their phone locks. 
%For respondents with partners, we also asked whether their partner's phone locked. 
Of those with a partner, 84.7\% (N = 341) said that their partner's phone locks~\footnote{8\% (N = 341) of respondents indicated that they did not know whether their partner's phone locked or not.}. We identify factors correlated with a participant locking their phone through a logistic regression model, summarized in Table~\ref{tab:lock_logreg} in the Appendix. 

We observe gendered differences with respect to phone locking habits: 82.1\% (N = 246) of women indicate that their phone locks, compared to 90.4\% (N = 197) of men, which is statistically significant ($\chi^2(8,460)=2.35, p=.019$). This implies that being Male increases the odds of locking one's phone by 134\% (95\% CI [1.17, 4.89]). In some locations, gendered cultural expectations limit women's freedom to lock their phones~\cite{Sambasivan}. While our data show that women lock their phones less than men, this does not appear to be \textit{because} of gender, but rather a reflection of an underlying gendered inequity in technological familiarity, which has been measured in other contexts~\footnote{While there is a digital gender gap among adults, educational research repeatedly shows none in K-12 students \cite{nogap1,nogap2}.}~\cite{digital_gendergap}. We discuss further demographic breakdown with respect to phone locking habits, as well as survey answers to authentication schemes in Appendix~\ref{app:authentication}.

%% file: table_mutual_access.tex
\begin{table}[]
\begin{tabular}{llllll}
                                                                                          &                          & \multicolumn{4}{l}{\begin{tabular}[c]{@{}l@{}}Partner Access to Respondent\end{tabular}} \\
                                                                                          & \multicolumn{1}{l|}{}    & \multicolumn{2}{l|}{Yes}                                & \multicolumn{2}{l}{No}           \\
\multirow{3}{*}{\begin{tabular}[c]{@{}l@{}}Respondent\\ Access to\\ Partner\end{tabular}} & \multicolumn{1}{l|}{}    & N               & \multicolumn{1}{l|}{\%}               & N             & \%               \\ \cline{2-6} 
                                                                                          & \multicolumn{1}{l|}{Yes} & 189             & \multicolumn{1}{l|}{0.554}            & 33            & 0.097            \\ \cline{2-6} 
                                                                                          & \multicolumn{1}{l|}{No}  & 39              & \multicolumn{1}{l|}{0.114}            & 80            & 0.235           
\end{tabular}
\caption{Summary of Couples' Access}
~\label{tab:mutual_access}

\end{table}

%% file: trust.tex
\section{Sharing and Trust}
\label{sec:trust}
In this section, we provide an overview of respondents' attitudes toward acceptable use of access to a partner's phone, perceptions of how common these usages are, and explanations offered for these responses. 
%Some people find all forms of access objectionable while others find them all desirable. Most people prefer something in the middle; device access is a binary, so one grants it but trusts that their partner will not abuse it. There is some consensus about what constitutes acceptable use of access, but disagreement is spread across all demographics and therefore seems to be a product of personal preferences.

%In open-ended questions at the end of each section, respondents were asked what additional context they could provide for their answers about the prevalence or desirability of certain behaviors. We find that regardless of someone's perspective on sharing, everyone understands their choice to be grounded in trust. In cases where access is granted but there are still constraints on appropriate use of access, boundaries must be negotiated, and everyone does so differently. Respondents emphasized that whatever the boundaries are in a given relationship, trust implies that they are respected, mutual, and consensual. In Section \ref{sec:betrayal} we discuss the propensity to, and justifications for, crossing these lines and betraying a partner's trust.

\input{6_1}

\subsection{Access, Boundaries, and Consent}
\label{sec:6_2}
The open-ended responses reveal the importance of context to the meaning of technology behaviors. In open-ended responses, 
some respondents say that they trust their partner, and therefore do not need access to their phone. 
Others say the trust in their relationship is complete, and therefore there is nothing to hide. 
One participant explains this succinctly: 
\begin{displayquote}
\textit{``Some relationships have a lot of trust where you don't want or need access to your partner's mobile device. In some relationships having access to your partner's mobile device and social media strengthens trust.'' ---Male, 29, `Serious' Relationship}
\end{displayquote}

Most of our respondents indicate that there should be some shared technology access in relationships. However, it is only acceptable with negotiated consent and \textit{trust.} %within established relationship parameters: access is viewed positively, snooping is not, and the line between the two is based on negotiated consent and \textit{trust.}
Participants also emphasize reciprocity of access, context and frequency of behaviors as factors in determining whether they are acceptable.

%\subsubsection{Importance of Privacy}
A sizable number of participants indicate that they either do not have access to their partner's phone or do not use it except for simple or benign tasks like navigating while their partner is driving. 
They explained they don't need to go looking through their partner's phone because they trust them: 
\begin{displayquote}
\textit{``My wife and I have a lot of trust with each other. I am unconcerned with what she does online or with her phone.
%I think phones are personal property, and like underwear, best used by their owner.
'' ---Male, 39, `Serious' Relationship}
\end{displayquote}
Many of these respondents view snooping behaviors as a \textit{symptom of distrust} in a relationship.
These participants were among the sizable number who feel that partners should have access to each others' phones and permission to look through anything they wish to access. 
%Participants explain this in different ways; 
Some feel this is the \textit{appropriate} state of affairs for a relationship,
%\begin{displayquote}
%\textit{``I feel all access at all times is acceptable'' ---Male, 72, `Serious' Relationship}
%\end{displayquote}
% \begin{displayquote}
% \textit{``I think it fair and just to have access to a partners phone
% '' ---Male, 39, `Serious' Relationship}
% \end{displayquote}
while others saw access as the default, with distrust being the only reason \textit{not} to grant it:
\begin{displayquote}
\textit{``I have a high level of trust with my partner so for us it's common sense to have pretty much total control of each others devices. It's convenient.'' ---Non-Binary, 29, `Serious'}
\end{displayquote}
As opposed to a default or \textit{path to trust,} 
some respondents feel that anything short of complete transparency would \textit{create distrust,} similar to findings by McDaniel et al~\cite{read_dms_ok}:
%\begin{displayquote}
%\textit{``Unless you['re] hiding something this should not be a problem'' ---Not provided, 56, single}
%\end{displayquote}
\begin{displayquote}
\textit{``I think it is good and healthy to have an open and honest relationship when it comes to seeing what your spouse has on their phone or social media account. If you are not comfortable with that, I would wonder what you were hiding.'' ---Female, 23, Relationship}
\end{displayquote}
% \begin{displayquote}
% \textit{``You can never trust someone 100\% unless you keep total tabs on them, and even then they will disappoint you.'' ---Female, 33, Relationship}
% \end{displayquote}
%\begin{displayquote}
%\textit{``I think every couple needs to have open communication and know what each other are doing online and who they are talking to'' ---Female, 36, Single}
%\end{displayquote}
%\begin{displayquote}
%\textit{``Marriages shouldn't have secrets. Having open communication and not hiding communication with others is important for a healthy marriage. I don't need to know every call, text, or DM, but if my spouse doesn't want me to, that's concerning.'' ---Female, 41, Single}
%\end{displayquote}
The largest group of respondents perceives the correct approach to be somewhere between transparency and privacy. 
Some say they ought to be allowed to look at anything, but never have,  \textit{because they trust} their partner. Psychological research has  identified this phenomenon as well~\cite{trust_mitigates_snooping}. 
%\begin{displayquote}
%\textit{``We're married. If I want to dig I can. We don't [though].'' ---Male, 43, `Serious' Relationship}
%\end{displayquote}
%\begin{displayquote}
%\textit{``I completely trust my partner. There is no secrecy between us. I know all his PINs/passwords and he would happily let me go through all of his messages, calls, etc. if I ever asked, but I've never felt the need to. Same vice versa.'' ---Female, 24 `Serious' Relationship}
%\end{displayquote}
% \begin{displayquote}
% \textit{``I don't need to check his phone because I trust him and vice versa. If he wants to look at my phone he is more than welcome to.'' ---Female, 41, `Serious' Relationship}
% \end{displayquote}

Others feel that access is needed to \textit{maintain trust,} but that it could be problematic if gained surreptitiously or through coercion, or used too often.
\begin{displayquote}
\textit{``I think pretty much all of it is normal but any could cause a problem. You can never fully trust anyone to behave in a way you deem appropriate.'' ---Female, 37, `Serious'}
\end{displayquote}

%\begin{displayquote}
%\textit{``I think trust is so important in [serious relationships] and that you should be able to have open access to each other's accounts. But secretly gaining access or watching EVERY little thing they do is super toxic.'' ---Female, 33, `Serious' Relationship}
%\end{displayquote}

%\begin{displayquote}
%\textit{``I think a lot of this behavior is fairly normal for modern relationships due to the stigma around social media and phones in general, but I do believe that almost all of it can be become a problem if it becomes constant or it's forced upon a partner.'' ---Male, 25, Single}
%\end{displayquote}
%Consent in the usage of access is a frequent theme in these responses. 
Some participants explain that their partner has access to their phone, but that everything within it is off-limits unless otherwise stated: %, and that they trust their partner not to overstep:  
\begin{displayquote}
\textit{``My partner and I will use each other's phones if it is easier. I would never look at anything that he wouldn't want me to and I trust that he wouldn't either.'' ---Female, 23, Relationship}
\end{displayquote}
% \begin{displayquote}
% \textit{``Having access is great, just don't abuse the power.'' ---Female, 61, Relationship}
% \end{displayquote}
Some participants emphasize that occasionally, transparency or confrontation is necessary to \textit{reinforce trust}, but that non-consensual access is never acceptable:
\begin{displayquote}
\textit{``I really don't look over or check things often, if I do it's because I'm feeling a negative type of way and just wanna make sure nothing is going on with her and someone else, each time I check, I get her consent for checking.'' ---Male, 20, Relationship}
\end{displayquote}
% \begin{displayquote}
% \textit{``Confronting your partner about stuff you've seen online is normal and helps build boundaries, but it gets toxic when reading messages.'' ---Non-Binary, 41, `Serious' Relationship}
% \end{displayquote}
% Participants hold diverse perspectives on appropriate levels and use of access; most feel that their preference is the appropriate manifestation of trust in a relationship.
% Some interpret trust to mean that one trusts their partner not to misbehave, and therefore does not need to check up on their behavior.
% Some view the ability to check up occasionally as necessary for the maintenance of trust.
% Some participants (implicitly) acknowledge a risk involved in giving someone blanket access but do so because they trust their partner. 
% Others grant access but establish boundaries, trusting their partner not to overstep. 
While participants hold diverse perspectives on appropriate levels and use of access, there is consensus that access, non-access, or mediated access should be reciprocal, consensual, not overused, and benevolent.

%% file: 6_1.tex
\input{table_attitudes}

\subsection{Attitudes Towards Sharing Practices}
\label{subsec:6_1}
For a series of behaviors, respondents were asked whether they thought that in the context of a romantic relationship, the behavior was \textit{very uncommon, somewhat uncommon, somewhat common, or very common.}
Then, respondents were asked about whether the same behaviors were \textit{bad/toxic, problematic, normal, or good/healthy.}
Some behaviors that were included are not specifically related to phones or digital technologies, such as sharing a bank account, and are included as baselines of measurement. 
Table \ref{tab:attitudes} provides a breakdown of responses.
Sample sizes range from 494 to 529.

\subsubsection{Acceptable Behaviors}
There is consensus about some behaviors being widely acceptable, with 80\% or more respondents deeming it `normal' or `good': sharing a streaming password, phone plan, bank account, or cloud account.
%While there are circumstances in which sharing financial accounts could be problematic, phone accounts can offer stepping stone access,
%\footnote{Leveraging access to one account to gain access to another, through password resets sent to an email, for example}
%as could cloud accounts similar to iCloud
%\footnote{iPhones are predicated on Apple accounts, which by default will back up text histories, location histories, and troves of other deeply personal data to iCloud. Further, iCloud charges a subscription fee for storage, and family plan rates incentivize sharing between users. There is a similar interaction between Android and Google Drive.}.
Additionally, 79.6\% think it is acceptable to share a phone PIN, and 75.3\% think it is acceptable to monitor a partner's social media behavior from one's own accounts.
The commonality and acceptability of sharing a phone PIN are borne out by our data, as we find that most people have access to their partner's phone, and if their partner's phone locks, they almost certainly have the PIN.
Many studies find that monitoring one's partners' public social media is common and not widely condemned~\cite{phone_conflict_autonomy,intercouple_agreement,attitude_norm_predict}, though other studies find that increased monitoring %of partners' social media 
is correlated with increased conflict over social media~\cite{twitter_breakup, SNS_conflict_notinfidelity,FBaddiction_leads_surveillance}.
%including one which found that those who thought of the behavior as normal and beneficial were more likely to do so. \cite{attitude_norm_predict}
%It's easy to understand thinking that monitoring a partner's social media from one's own account is acceptable---it is, after all, on the open web.
%However, many of these same studies find that increased monitoring of partners' social media is correlated with increased conflict over social media~\cite{twitter_breakup, SNS_conflict_notinfidelity,FBaddiction_leads_surveillance}.
Perhaps it is then unsurprising that while only 45.7\% of respondents feel confronting a partner about something they'd done on social media is acceptable, 70.8\% say it is common. 

Some behaviors are undesirable (fewer than 20\% of respondents indicating it is acceptable), including active snooping behaviors: logging into a partner's social media, reading their DMs, posting/changing something, checking search, location, or call history, and reading texts. Participants explain that these are different from monitoring from one's own account: 
\begin{displayquote}
\textit{``I don't consider it to be bad or toxic to look at my partner's social media profiles. I do however believe it's taking a bad turn when you're actually logging in and reading messages.''---Female, 33, `Serious' Relationship}
\end{displayquote}
%We investigate individual boundaries and mental frameworks in \ref{sec:6_2}

The least acceptable behavior is \textit{``install an app that tracks or monitors my partner, or otherwise relays data about them and their phone back to me''}---`Non-Mutual Tracking App' in Table \ref{tab:attitudes} ---representing apps often marketed as child tracking apps, but which are essentially spyware~\cite{rahul}.
Contrastingly, \textit{``Use a couples' tracking app''}, representing apps designed for \textit{mutual} tracking---``Mutual Tracking App'' in Table \ref{tab:attitudes} ---has significantly more approval ($t(994)=9.37, p<.0001$).
Most respondents (73.7\%) still find this behavior concerning, but the difference underscores our finding that norms of acceptability are context-dependent;
%the acceptability of behaviors depends on the context, meaning, and motives for their use. M
mutual and consensual behaviors are more acceptable.

%Further research could investigate whether there is a relationship between comfort with this behavior and issues of jealousy and control within relationships; while some of these apps are benign, and are effectively just games or chat apps, some couple tracker apps may include logs of the partner's text communication, location, or other information\footnote{\url{https://coupletracker.com/}}. There is a slippery slope from well-intentioned couples' apps to what amounts to spyware; the role of mutual and unidirectional tracking apps in intimate partner violence is well documented~\cite{IPS_narratives, rahul}.

While some behaviors garner general condemnation, it is important to keep in mind that for every single behavior, some people said the behavior was good, and therefore in discussing betrayals in Section \ref{sec:betrayal} we intentionally examine only \textit{non-consensual} interactions.
The need for consent in digital interactions is further emphasized by the sharp difference in perceptions of acceptability between \textit{ ``check location history''}---deeply undesirable---and \textit{ ``share location on an ongoing basis''}---no consensus. 
Though either yields the same information, one is unidirectional and surreptitious, while the other is mutual and consensual.

%There were some areas without consensus though, respondents were split on whether it was acceptable to share social media passwords, email passwords, location on an ongoing basis, or give a partner biometric access. 
%There are demographic distinctions regarding these behaviors as discussed in \ref{subsub:attdemo}, but none had either strong support (> 80\% acceptable) or condemnation (< 20\% acceptable) for any demographic group. 

There are notable differences between perceptions of the acceptability and prevalence of some sharing behaviors.
For general sharing behaviors\footnote{Sharing PIN, biometric access, social media, email, and streaming passwords, bank accounts, and phone plans.}, significantly more people think they are acceptable than think they are common ($t(9254)=16.46, p<.0001$.
For snooping behaviors~\footnote{Checking search, location, or call history, reading texts or DMs, logging into or taking action within partner's social media},
significantly more people think they are common than think they are acceptable ($t(9120)=-19.98, p<.0001$).
This is also the case for confronting a partner about social media ($t(1011)=-8.38, p<.0001$).
Finally, there is a perception that usage of unidirectional tracking apps is significantly more common than acceptable ($t(1010)=-4.53, p<.0001$)~\footnote{The only behaviors without significant differences were: ongoing location ($t(1026)=-.01, p=.989$), and mutual tracking app ($t(1006)=.33, p=.742$).}.

These results demonstrate a pessimism in responses:
\begin{displayquote}
\textit{``I think a lot of these actions are common, and most of them are toxic. Keeping tabs on your partner, or snooping through their stuff is not ok.''---Male, 26 `Serious' Relationship}
\end{displayquote}
%
%\begin{displayquote}
%\textit{``I'd like to hope that there are more healthy relationships than unhealthy relationships, but I feel like I hear more about people who are jealous or suspicious rather than people in healthy loving relationships''---Non-Binary, 25, Single}
%\end{displayquote}
However, respondents also note that they aren't sure whether their perception of frequency is accurate, as it comes in large part from media, particularly social media and television: 
\begin{displayquote}
\textit{``I see couples on TV and on crime shows spying on each other all the time but I don't know how common it is in real life.''---Female, 58, `Serious' relationship}
\end{displayquote}

\begin{displayquote}
\textit{``I think people talk about looking at each other's phones a lot on social media but I don't think it's super common unless they don't trust each other.''---Female, 20, Relationship}
\end{displayquote}
%
%\begin{displayquote}
%\textit{``A lot of memes and social media seem to normalize snooping through your partner's phone or social media accounts.''---Female, 34, In a relationship}
%\end{displayquote}
This doubt is justified; studies show that television portrayals of couples' interactions do not reflect real-life behavior\cite{TV_couples_inaccurate},
and that depictions of security behaviors create problematic misconceptions~\cite{security_media}.
%It is established that perceiving a behavior to be accepted or common leads people to engage in it~\cite{attitude_norm_predict}, so future work could investigate the role of media portrayals, and the relationship between social media algorithms which incentivize provocative content \cite{algoprobs}, the tendency to misrepresent oneself on social media\cite{faceyfacebook}, and perceptions about snooping behaviors.

\subsubsection{Demographic Considerations}
\label{subsub:attdemo}
There were few gender differences in perceptions about what behaviors are acceptable, with the exception of confronting a partner about social media: 49\% of women think it is acceptable compared to 39\% of men ($\chi^2(1, 472)=5.76, p=.016$). 
%There is less agreement about how common behaviors are; 20\% of women think using a couples' tracker is common, compared to 29\% of men. This difference is not significant after statistical corrections ($\chi^2(1, 480)=4.50, p=.034, \alpha_{BH}=.027$)~\footnote{Men also thought it was more common to share a cloud account ($\chi^2(1, 493)=9.67, p=.002$).}. 
Women are significantly more likely to think that most snooping behaviors are common, including checking call history on the device ($\chi^2(1, 487)=8.99, p=.003$) or the phone bill ($\chi^2(1, 486)=5.48, p=.019$), reading texts ($\chi^2(1, 488)=13.02, p=.0003$), as well as confronting one's partner about social media ($\chi^2(1, 486)=14.99, p=.0001$).

Previous research has also found that men are more likely to actually attempt to log into a partner's social media account \cite{men_login_more},
and women are more likely to engage in snooping behaviors~\cite{trust_mitigates_snooping, intercouple_agreement}.
Furthermore, these patterns emerge in Norton's data: significantly more women (31\%) than men (27\%) acknowledged having \textit{``checked their partner's phone to view text messages, phone calls, DMs, emails or photos''}, while significantly more men (15\%) than women (6\%) had \textit{``used an app to monitor their partner's text messages, phone calls, DMs, emails, or photos''}.
%While both of these behaviors may be used to monitor a partner's activities, the context and impact may be qualitatively as well as quantitatively different. 
%Similarly, men in Symantec's survey were significantly more likely to have \textit{``tracked their partner's location via a location sharing app like Google Maps or Find My Friends''} or \textit{``tracked their physical activity via their phone or health app''}. 
These findings suggest that men are more likely than women to engage in (or be cognizant of others engaging in) active, ongoing surveillance, while women are more likely to engage in (or perceive higher prevalence of) opportunistic snooping. The Norton data supports this: 29\% of male respondents said they were familiar with ``stalkerware''
\footnote{Question: \textit{How familiar are you with stalkerware? Stalkerware is software that enables someone to monitor the activities on another person' device without that person's consent or knowledge.}}, while only 15\% of women were.% This difference is statistically significant.

Our findings indicate generational differences in perceived acceptability of sharing behaviors.  There is only generational consensus about the acceptability of sharing a PIN ($\chi^2(3, 506)=5.49, p=.139$), ongoing location ($\chi^2(3, 506)=.76, p=.859$), cloud account ($\chi^2(3, 506)=7.02, p=.071$), and phone plan ($\chi^2(3, 506)=3.71, p=.295$)~\footnote{$\chi^2(3, 506)$ tests were performed, all other questions $p-value < .05$}.
Much of the disagreement stems from Gen Z; these respondents are more comfortable with sharing biometric access (75\% thought it was acceptable, compared to 60\% of other respondents), but less comfortable with invasive and snooping behaviors.
For example, 1\% of those in Gen Z think unidirectional tracking apps were acceptable compared to 8\% of other respondents, and 6.5\% say checking a partner's search history is acceptable, compared to 20.5\% of participants in other generations. 
This trend persists for checking location and call history, and reading texts and DMs.
Baby Boomers are the other generation which differs from the rest: these respondents are less accepting of monitoring social media (54\% compared to 77\% of other respondents, $\chi^2(3, 506)=13.13, p=.004$) and confrontations about social media (23\% compared to 47\% of other respondents, $\chi^2(3, 506)=13.82, p=.003$)~\footnote{Baby Boomers were also significantly more disapproving about sharing streaming passwords (78\% approve compared to 97\% of other respondents, $\chi^2(3, 506)=29.59, p<.0001$).}.
Generations that were more accepting of a behavior generally thought it was more common as well, though not all of these differences were significant.

Respondents whose relationship status are single, `serious', and `less serious' are in agreement on social media, with no significant difference in acceptance of monitoring ($\chi^2(2, 506)=2.03, p=.362$), logging into ($\chi^2(2, 506)=5.80, p=.055$), or confronting a partner about social media ($\chi^2(2, 506)=.61, p=.737$).
Respondents in `serious' relationships are significantly more likely to approve of all other behaviors than those in `less serious' relationships.
Single participants are significantly less accepting of credential sharing
\footnote{Sharing PIN ($\chi^2(2, 506)=30.65, p<.0001$) or biometric access ($\chi^2(2, 506)=27.71, p<.0001$), social media ($\chi^2(2, 506)=26.21, p<.0001$) and email passwords ($\chi^2(2, 506)=31.94, p<.0001$), and cloud accounts ($\chi^2(2, 506)=8.69, p=.013$)}
than those in relationships. 
These findings are complicated by age: while Gen Z was the generation with the largest population of singles, they were not significantly more likely to disapprove of any of these behaviors than other generations.

%% file: table_attitudes.tex
\begin{table*}[h]
\centering
\begin{threeparttable}
\begin{tabular}{l|l|lll}
\begin{tabular}[c]{@{}l@{}}\textit{In a romantic relationship,}\\ \textit{this behavior is:}\end{tabular} & \begin{tabular}[c]{@{}l@{}}Somewhat\\ or Very\\ Common\end{tabular} & \begin{tabular}[c]{@{}l@{}}Good/\\ Healthy\end{tabular} & Normal & \begin{tabular}[c]{@{}l@{}}Bad/Toxic or\\ Problematic\end{tabular} \\ \hline
Sharing:                                                                                               &                                                                     &                                                         &        &                                                                    \\ \hline
Phone PIN                                                                                              & 0.589                                                               & 0.565\tnote{\dag}                                                   & 0.231  & 0.204                                                              \\
Biometric                                                                                              & 0.39\tnote{\ddag}                                                                & 0.487\tnote{\dag\ddag}                                                   & 0.164  & 0.349                                                              \\
Social Media Password                                                                                  & 0.377\tnote{\ddag}                                                               & 0.414\tnote{\dag}                                                   & 0.142  & 0.444                                                              \\
Email Password                                                                                         & 0.334\tnote{\ddag}                                                               & 0.425\tnote{\dag}                                                   & 0.16   & 0.415                                                              \\
Streaming Password                                                                                     & 0.915\tnote{\ddag}                                                               & 0.24                                                    & 0.712\tnote{\ddag}  & 0.048                                                              \\
Ongoing Location                                                                                       & 0.581                                                               & 0.409                                                   & 0.172  & 0.419                                                              \\
Cloud Account                                                                                          & 0.485\tnote{*\ddag}                                                               & 0.462\tnote{\dag}                                                   & 0.351  & 0.187                                                              \\
Bank Account                                                                                           & 0.746\tnote{\ddag}                                                               & 0.508\tnote{\dag}                                                   & 0.345  & 0.147                                                              \\
Phone Plan                                                                                             & 0.856\tnote{\ddag}                                                               & 0.352                                                   & 0.57\tnote{\dag}   & 0.088                                                              \\ \hline
Using:                                                                                                 &                                                                     &                                                         &        &                                                                    \\ \hline
Mutual Tracking App                                                                                    & 0.253\tnote{\ddag}                                                               & 0.19                                                    & 0.073  & 0.737\tnote{\dag\ddag}                                                              \\
Non-Mutual Tracking App                                                                                & 0.138\tnote{\ddag}                                                               & 0.039                                                   & 0.016  & 0.945\tnote{\ddag}                                                              \\ \hline
Checking Partner's:                                                                                    &                                                                     &                                                         &        &                                                                    \\ \hline
Search History                                                                                         & 0.432                                                               & 0.132                                                   & 0.022  & 0.846\tnote{\dag\ddag}                                                              \\
Location History                                                                                       & 0.342                                                               & 0.122                                                   & 0.026  & 0.852\tnote{\dag\ddag}                                                              \\
Call History (On Phone)                                                                                & 0.395\tnote{*\ddag}                                                               & 0.114                                                   & 0.03   & 0.856\tnote{\dag\ddag}                                                              \\
Call History (Phone Bill)                                                                              & 0.336\tnote{*\ddag}                                                               & 0.15                                                    & 0.034  & 0.816\tnote{\dag\ddag}                                                              \\
Text History                                                                                           & 0.462\tnote{*\ddag}                                                               & 0.1                                                     & 0.03   & 0.87\tnote{\ddag}                                                               \\ \hline
Partner's Social Media:                                                                                &                                                                     &                                                         &        &                                                                    \\ \hline
Monitoring From Own                                                                                    & 0.849\tnote{\ddag}                                                               & 0.465\tnote{\ddag}                                                   & 0.288  & 0.247                                                              \\
Logging Into Account                                                                                   & 0.312                                                               & 0.162                                                   & 0.049  & 0.789\tnote{\ddag}                                                              \\
Reading DMs                                                                                            & 0.309\tnote{\ddag}                                                               & 0.079                                                   & 0.037  & 0.884\tnote{\dag\ddag}                                                              \\
Making A Post Or Change                                                                                & 0.188\tnote{\ddag}                                                               & 0.067                                                   & 0.018  & 0.915\tnote{\dag\ddag}                                                              \\
\begin{tabular}[c]{@{}l@{}}Confronting Partner About\\ Social Media Activity\end{tabular}              & 0.708\tnote{*}                                                               & 0.36                                                    & 0.1    & 0.54\tnote{*\ddag}                                                              
\end{tabular}
\begin{tablenotes}
\footnotesize

\item[*] Significant Differences between Genders at $\alpha=0.05$ with Benjamini-Hochberg Corrections 
\item[\dag] Significant Differences by Relationship Status at $\alpha=0.05$ with Benjamini-Hochberg Corrections 
\item[\ddag] Significant Difference across Generations at $\alpha=0.05$ with Benjamini-Hochberg Corrections 
\end{tablenotes}
\end{threeparttable}
\caption{Respondents Perceptions of Commonality and Acceptability of Behaviors}
~\label{tab:attitudes}

\end{table*}

%% file: betrayal.tex
\section{Betrayals and Justifications}
\label{sec:betrayal}

%We demonstrate that different levels of access and transparency are desirable or comfortable for different people.
%However, we find that regardless of individual preferences, one large point of consensus is that access should be mutual and consensual. 
%A large point of consensus amongst our participants is that access should be mutual and consensual. 
Here, we investigate \textit{non-consensual access}.
%drawing on our own data and text responses, as well as Symantec's data, and prior work.

%We discuss the frequency with which non-consensual access occurs, as well as the rationales and justifications provided for doing so. 
%We find that concerns about infidelity are \textit{the dominant} explanation given for overstepping boundaries, which is supported by prior work demonstrating that social media usage can lead to increased jealousy in relationships, and thus lead to snooping and confrontation. \cite{}
%These same concerns have been shown to be a driving force in adoption of spyware and other partner surveillance, \cite{} and we also discuss the relationship between snooping and abuse.

%- our numbers and symantec numbers on how often this stuff happens

%- our numbers in context of whether people think this is ok

%- our free text responses
%- symantec's why questions!!

\subsection{Frequency of Overstepping}
% A number of social and computer science studies investigate the prevalence of digital snooping behaviors. However, this research does not always make clear whether the behaviors were surreptitious and/or non-consensual. 
% As we demonstrate in \ref{subsec:6_1}, the meaning of behaviors that some researchers presume to be invasive, such as reading a partner's texts, are actually context-dependent. 
% Norton's data explicitly focuses on access that is non-consensual, asking \textit{``Which of the following, if any, have you ever done to a current or former significant other/romantic partner without their knowledge or consent?''}
% Reed et. al.'s research with college students~\cite{addl_snoop_ipv_stats} also investigates whether behaviors are consensual.

%\peri{Consider moving this to a table}

Norton finds that 46\% of people have `online stalked' a current or former partner, a figure which is not surprising given findings in other work that `Facebook stalking' is fairly common~\cite{FB_maintenance_jealousy,phone_conflict_autonomy}.
There is legitimate debate about whether consent is needed to monitor a partner's social media from one's own account and devices; 29\% of our respondents say doing so is healthy, and 45\% say it's normal. 
Norton's respondents are also ambivalent about this type of monitoring; 35\% say they don't care if they are being online stalked by a current or former partner, and 30\% say it is harmless. Significantly more men felt this way than women.

More active forms of monitoring are less common among Norton's respondents: 
9\% had ``catfished'' their partner,
%~\footnote{Creating a fake profile to interact with their partner or view content they may not otherwise see.}
8\% had tracked their physical activity, and 10\% had used an app to monitor a partner with other people.
Men were significantly more likely than women to do the last two.
%\molly{can we break the items with significant differences out by sex? eg Significantly more men felt this way than women.}
Norton finds that 25\% of those under age 35\footnote{This is the granularity at which the age data is available.}
%16\% of respondents 
have tracked their partner's location non-consensually, 
while Reed et al. find that 36\% of college students have done so~\cite{addl_snoop_ipv_stats}.
Norton finds men significantly more likely to track a partner's location without consent (in general, and among young respondents), while Reed et al. find no gender differences~\cite{addl_snoop_ipv_stats}. %in non-consensual location monitoring.
Conversely, Norton finds that 
%29\% of respondents (
42\% of respondents younger than 35 have accessed a partner's phone to look at texts, calls, DMs, emails, or photos without consent, reporting no sex differences. 

Since our study was concerned with general sharing behavior, we did not explicitly ask about non-consensual behavior. However, we did ask participants who had access to their partners' phones how they used it. %Our questions about participants' practices did not align exactly with our questions about social norms.
%, instead focusing on behaviors which were predicated on access to a partner's device. 
%Thus, we did not ask about credential sharing, or monitoring of social media from one's own accounts. 
%We also asked some more granular questions about installing apps and changing settings. 
For items that appear in both behavior perceptions and actual behavior sections
\footnote{Reading texts, logging into partner's social media, reading DMs, taking other action within social media, checking search history or location history, and installing unidirectional tracking apps.}, respondents who engage in the behavior are significantly 
\footnote{2-sided, 2-sample t-tests, those who had engaged compared to the remainder, all p-values < 0.05}
more likely to find the behavior acceptable and common than the remainder of the sample. 
%This follows from findings that those who view Facebook stalking one's partner as normal and acceptable were more likely to engage in it~\cite{attitude_norm_predict}.

\subsection{Justifications for Overstepping}
Respondents offer varying rationales for overstepping their partner's boundaries. 
We find that a majority of responses discuss infidelity, though participants also discuss other specific concerns leading to snooping, as well as generic justifications. %, including building or repairing trust, and a sense that there was no harm in doing so. 
%Previous work has also investigated when these behaviors are perceived to be justified. \cite{}

\subsubsection{Trust Building}
As discussed in \ref{subsec:6_1}, %and previous research~\cite{infidelity_perception_monitoring}, 
transparency can function to mitigate normative mistrust in some relationship contexts.
Some participants explained that snooping offers this transparency, helping to rebuild trust after some form of betrayal: 
\begin{displayquote}
\textit{``When you lose trust in someone who has hurt you, you are always snooping till the trust is earned back'' ---Female, 20, Relationship}
\end{displayquote}
Others indicate that snooping without an inciting betrayal helped them to establish trust in their partner, leading them to stop snooping. 
The role of trust in mitigating snooping has been previously documented~\cite{trust_mitigates_snooping}. 
\begin{displayquote}
\textit{``I'm not proud to say that I've looked at some of my boyfriend's texts out of curiosity to see who he was taking to. I don't do this anymore and I found nothing interesting'' ---Female, 20, Relationship}
\end{displayquote}
While Norton did not ask about infidelity, 27\% of respondents who acknowledge snooping on their partner in some capacity said it was because \textit{``I didn't trust them,''} and 34\% said they had done so because \textit{``I suspected they were up to no good.''} Women were significantly more likely to cite distrust, while there was no gender difference on the latter answer.

\subsubsection{Behavioral Concerns}
Responses offering specific reasons for snooping focused largely on infidelity, but there were some notable exceptions. Some participants discuss concern for their partners' safety or well-being, or a need to verify they were not engaging in problematic behaviors:
\begin{displayquote}
\textit{``I have a tracker on his phone that allows me to see where he is at all times. I also have an app on my phone that allows me to see what he is doing on his phone. He has a drinking problem and I make sure he is not somewhere that has alcohol.'' ---Female, 25, `Serious' Relationship}
\end{displayquote}
Norton finds that 24\% of respondents had snooped on their partner because \textit{``I wanted to make sure they were safe, physically and/or mentally.''} A further 30\% said \textit{``I wanted to know where they were.''} Men were significantly more likely to give both answers. It is unclear \textit{why} these participants want to know where their partner is, or whether concern for their safety is warranted. 
%The quote above provides an example of incredibly invasive behavior with an ostensibly legitimate justification.

Several respondents cite their partners' pornography consumption as a reason they snoop. %Research has shown that there is significant disagreement between men and women about whether consuming online pornography is acceptable in relationships, with women being significantly less accepting \cite{porn_prevents_infidelity,intercouple_agreement}.
%\begin{displayquote}
%\textit{``I make sure he is not visiting pornography sites that he likes to.'' ---Female, 55, `Serious'}
%\end{displayquote}
%\begin{displayquote}
%\textit{``He has had a pornography problem in the past so I check on him.'' ---Female, 21, `Serious'}
%\end{displayquote}
%
For some people, consuming pornography can equate to a kind of mental infidelity~\cite{doi:10.1080/10720162.2012.658344}: 
\begin{displayquote}
\textit{``My husband is the first person I've dated that I trust. Lots of cheating, lots of looking at porn secretly, lots of dming other people.'' ---Female, 24, `Serious' Relationship}
\end{displayquote}
%However, research has shown that consuming pornography has a deterrent effect on infidelity \cite{porn_prevents_infidelity}. 

\subsubsection{Infidelity}
Infidelity is the dominant reason offered for violating agreed boundaries; those who had snooped cited concerns about cheating, with many respondents attributing infidelity to social media (though research is inconclusive as to whether this perception is accurate~\cite{SNS_conflict_notinfidelity,SNSaddiction}). 
%\begin{displayquote}
%\textit{``I think social media is harming relationships. I found out my husband messages with several old girlfriends but he got very upset when I did this.'' ---Female, 59, `Serious'}
%\end{displayquote}
%\begin{displayquote}
%\textit{``I believe that social media is a tool for facilitating infidelity and society is becoming more aware of this danger.'' ---Male, 36, Single}
%\end{displayquote}
%Research is inconclusive as to whether this perception is accurate; some studies have shown that social media use is not related to infidelity~\cite{SNS_conflict_notinfidelity}, while others have found that social media addiction is related to `social media infidelity'~\cite{SNSaddiction}. As mentioned in \ref{subsec:6_1}, further work could investigate the role of social media in shaping this perception, not just in facilitating infidelity.
People indicate that suspicion of infidelity is a valid justification for snooping: 
\begin{displayquote}
\textit{``It would be an invasion of privacy if your partner logged into your social media and checked your messages for no reason. However, if one partner suspects the other of cheating, I think it is bad (but justifiable) to do it.'' ---Male, 26, Single}
\end{displayquote}
These sentiments are echoed in Nortons's data: 31\% of respondents -- significantly more men than women -- indicate that \textit{``Online stalking is okay if one or both partners have cheated or are suspected of cheating.''}
Many respondents offer nuanced, contextualized explanations for this; a frequent sentiment is that it is justified to snoop if you have \textit{reasonable} suspicions that your partner is cheating, but that being too quick to become suspicious is controlling or toxic.
%\begin{displayquote}
%\textit{``If you think your partner is being unfaithful you have every right to snoop. But if you're snooping because you have a guilty conscience that is a problem.'' ---Female, 50, `Serious'}
%\end{displayquote}
%\begin{displayquote}
%\textit{``If there's reason for suspicion, most of those actions are warranted. If a partner's suspicious from the start, that is controlling behavior.'' ---Female, 27, Relationship}
%\end{displayquote}
%

While these participants are justifying non-consensual access to a partner's device, they reinforce our understanding that all norms are contextual; snooping isn't acceptable \textit{unless} there's a suspicion of cheating, and even then, \textit{only} if that suspicion is legitimate. 
Technology behaviors that are positive in healthy relationships can be abusive in coercive and controlling relationships \cite{tech_ipv_contextual, Dragiewicz2021, DragiewiczMumPW}, and it appears that similarly, behaviors which are never desirable can be acceptable or justifiable in some contexts but not others.

%% file: discussion.tex
\section{Discussion}
The survey data reported here demonstrate that individuals grant romantic partners access to their smartphones more often than not, with little variation across demographic categories. 
Previous work indicates that sharing is something people do because it is convenient, but are only willing to do with those they trust~\cite{caring_sharing, sunny_sharing}.
Our data demonstrate that while access is convenient, trust may be the reason for access. 

Our data, the Norton survey, and the extant research show that there is no singular consensus about what constitutes desirable or undesirable sharing of partners' phones. 
In some relationships, complete transparency is consensual and used to establish and maintain trust. However, many participants indicate that they have boundaries regarding what their partner can do on their phone, and would feel violated if they overstepped. 
The primary difference separating the first circumstance from the second is the consent of both parties.
While trust is important in relationships, we find---as have many others---that those with access to partners' phones and rules about its usage often violate these boundaries, with potentially serious implications. 
As such, we suggest making some of these interpersonal, often implicit, boundaries technologically enforceable.

\subsection{Design Recommendations}
%\peri{this ordering is weird can fix later}

%\peri{way too long remove like first 1.5 paragraphs}

Smartphones are largely designed to be single-user devices, as evidenced by the lack of secondary authentication for credentialed services;
%as evidenced by a recent shift to one's phone as a primary token for validating identity---previously this would have been an account. 
%The usage of phones as both an authentication token and an access-controlled user interface is problematic. 
%Two factor authentication is largely predicated on phones; 
%much is still via SMS, though services are adopting OTP-generating apps like Google Authenticator, and Google and Apple use direct push notifications to verify account activity (on their respective operating systems). 
%Similarly, organizational users are adopting systems like Duo, which are also linked directly to the device, not the number.
in having access to a person's phone, one likely also can access many of their accounts and even reset passwords to others. 
%
%This is the security wonk's version of the problem. 
%For many people, the concerns surrounding someone poking around their phone are more likely to do with interpersonal harms than financial ones: 
%the same texts that hold OTPs hold private conversations. 
And yet, many people find it convenient and desirable to give their partner access to their device, trusting their partner not to abuse that access.
There are both security and privacy concerns with a partner reading texts and emails, but there are also incentives to give a partner device access. The solution should be to make it possible \textit{to prevent} access to these communications, while still granting access to the phone. Many circumstances make it impossible or undesirable to avoid sharing~\cite{Sambasivan, ACCAN,eSafetykids2020,Douglasetal2019,Woodlock_McKenzie_Western_Harris_2020, DragiewiczMumPW}, and \textit{most users} share their devices for a range of legitimate reasons.
%Currently, the dominant attitude of the security sector seems to be that one should simply not share their device. However, this attitude is incompatible with users' behavior and needs. Dominant security models fail to account for circumstances that make it impossible or undesirable to avoid sharing, \cite{Sambasivan, ACCAN,eSafetykids2020,Douglasetal2019,Woodlock_McKenzie_Western_Harris_2020, DragiewiczMumPW} and the fact that \textit{most users} share their devices for a range of legitimate reasons.

The makers of operating systems could consider adding (or promoting) multiple profiles to devices with different permission levels, as one would use a `guest' account on a PC or a different user account in multi-user smarthomes~\cite{zeng2019understanding,geeng2019s,mare2020smart, sikder2020kratos}. 
One imagines this alternate profile could have access to specific apps---music, navigation, games---without having access to the device owner's communications, social media, or other sensitive information. 
The device owner could ideally select which apps are shared, potentially creating a generic `guest' profile for anyone who needs it and a dedicated profile for those whom they permit access to more private things. 
Through such a system, access could be customized to grant multiple users access to specific services, with adjustable settings to account for changes in relationships. While multi-user accounts could raise an issue of coercion, in relationships with differential power dynamics such as parent-child~\cite{ParentsT7:online} or intimate partner violence~\cite{stalker_paradise}, the individual with more power often already has access to the other person's phone. 

Alternatively, more app makers could adopt session-level authentication. 
Many banking and financial apps require authentication with each session, either a dedicated username and password or default to the device's authentication after the first login.  Vault apps, which provide more private storage, require PIN authentication every time the app is opened. Social media, email, and other communication apps could make this change. On Android, users can select their default SMS app, and there are options like Signal which encourage users to require authentication for each session. Apple could consider adding this feature to iMessage. Future work is needed to see how to design for the usability trade-off of additional authentication.
If more app makers allowed and encouraged additional authentication, %---seamlessly, through biometrics---
this could address many individuals' concerns; in reality, it's only a few specific behaviors people worry about in terms of snooping.

While granting a partner biometric access is significantly less common than PIN-based access, the changes above would be mitigated by such access. 
%Operating systems could consider restricting biometrics to one face or one fingerprint. 
%This is reasonable security advice broadly: even when both are on the same hand, enrolling two fingerprints on a device significantly increases the likelihood of someone else being able to unlock it. \cite{} 
%Further, this change might allow operating systems makers to lean further into biometrics. 
Currently, the PIN is authoritative on both mobile operating systems; it's the encryption key that unlocks the hard drive, and if a sensitive settings change requires authentication, it can't be done with biometrics.
The rationale here is that biometrics can fail or be fooled, and are not infinitely reliable. However, while a machine will never misinterpret a PIN, they are far more easily abused. 
Operating systems could seek to use biometrics for authorizing everything short of encryption, and default to the underlying Apple or Google password when biometrics fail~\footnote{The authors acknowledge that there are highly adversarial situations where it is necessary to have only knowledge-based authentication, and would suggest that users with this threat model decline to enable biometrics and use passwords.}.
Alternatively, dual PINs could be implemented for tiered access. %Whatever the mode of authentication, 
Also, making a log of access times and modalities available to users could be useful as evidence in intimate partner abuse cases.

\subsection{Future Work}
%There is much room for future work evaluating technology users' privacy concerns and how the tech industry can best design systems to address them. Future work is needed to provide a concrete implementation of the design suggestions offered here, or offer improvements upon them. It will also be critical to understand the usability of these changes, and the impact they will have on the interpersonal and relationship issues being studied here; conflicts might arise in the process of making implicit boundaries explicit and enforceable, other conflicts might be avoided.

While this paper provides important new data to help us understand phone-sharing norms, further research could further investigate the extent to which these norms are gendered, and whether there is consensus in peer groups. 
Many respondents discussed the way their perceptions of specific behaviors were influenced by popular media; %, indicating that they thought the true prevalence of infidelity and snooping was lower than portrayed in these media.
research should investigate whether the true prevalence of infidelity and snooping is the same as portrayed in media, and the role media plays in shaping norms. %, particularly when one considers what is understood about social media algorithms---inflammatory content performs very well, so monetized accounts have an incentive to skew towards the extreme~\cite{algoprobs}.
And studying relationship implications of these norms, i.e., context, meaning, motives, or outcomes of behaviors for those involved beyond college student samples~\cite{SNS_conflict_notinfidelity,SNSaddiction,twitter_breakup}, is important to be able to make concrete security policy and recommendations.
%Just as norms for technology sharing are poorly understood, we know little about the relationship implications of these norms. The interaction between social media usage and snooping and negative relationship outcomes has been studied, especially with college student samples~\cite{SNS_conflict_notinfidelity,SNSaddiction,twitter_breakup}, but this research infrequently asks about the context, meaning, motives, or outcomes of behaviors for those involved. In addition, research on consensual access to intimates' devices is sparse. Research on people's actual behavior, including sharing, is prerequisite to developing adequate security policy and practice.
Finally, further work is needed to study how the existence of sharing behaviors impacts technology-facilitated abuse. %Closed-ended survey questions that simply ask about the presence or absence of specific behaviors are likely inadequate because they may produce data that is difficult to interpret without context and impact measures. Our study demonstrates that the meaning of discrete behaviors is largely context-dependent. 
While our study shows that even highly sensitive sharing, such as access to bank, cloud, and mobile accounts are common in relationships, they pose an extreme danger in the context of coercive control~\cite{Dragiewicz2021, DragiewiczMumPW}. %Research with abuse survivors and other users can help build understanding of the potential implications of the design changes we name above.
%Some people feel total access and moderate snooping as normative---perhaps these relationships are problematic, but it seems paternalistic to assume that a behavior can never be engaged in by free, mutual consent. 

%% file: conclusion.tex
\section{Conclusion}

In this work, we analyze two large-scale surveys, one our own survey (\textit{N} = 531) and the other from Norton (\textit{N} = 2,050), to examine shared smartphone access with romantic partners as a function of trust in relationships. We demonstrate that shared smartphone access is not always negative in relationships, %Much research may focus on invasions of privacy, snooping, and the relationship these have with abuse--- identifying behaviors that are presumed to be negative, and seeing privacy as the solution. However, similar to others~\cite{sunny_sharing, caring_sharing} we find that people in healthy relationships share access to their phones with their partner. 
and that a majority do so and consider this desirable. %It is therefore prudent to think about smartphone access and sharing not in the context of \textit{if}, but \textit{how, when, and why.}
While there are few points of clear consensus from our participants about what constitutes appropriate sharing in relationships, %with some participants favoring complete privacy from their partner and others desiring total transparency and access to their accounts, 
we find consensus that access behaviors need to be reciprocal and consensual to be positive and that most relationships involve access parity.
%While security experts and social scientists might favor privacy or transparency, neither can be viewed as inherently positive or negative in intimate contexts.
%In either case, users view these choices as a product of trust; either trust that negates a need for access, or trust that mitigates concerns about granting it.  Many partners provide access but set boundaries, relying on trust to enforce them. 
We also demonstrate, like prior work~\cite{rahul,addl_snoop_ipv_stats,distrust_leads_snooping}, that despite setting trust-enforced boundaries around access, partners over-step boundaries and abuse the access they have been given. %, snooping, monitoring, and otherwise betraying the trust that was supposed to enforce the boundary.
We make design recommendations for mobile applications and operating systems to have settings for %should be made enforceable through changes to mobile applications and operating systems, which would provide users with settings choices that 
reflecting the nuance of boundaries in relationships, and may also help protect those with abusive partners.

%We hope that in providing this detailed analysis of practices, attitudes, norms, and rationales, we are able to contribute critical context to a discussion about designing consumer systems so that they protect users from intimate threats, both severe and slight. 

%% file: appendix.tex
\section*{Appendix}

\section*{Phone Locking and Demographic Considerations}~\label{app:authentication}
\input{table_locking_logreg}

Table \ref{tab:rel_age_lock} provides a breakdown of phone locking behavior by age and relationship status. There are significant generational differences in rates of phone locking, with genz significantly more likely to lock their phones ($\chi^2(8,460)=2.02, p=.044$). Generation Z are 139\% more likely to lock their phones than other generations (95\% CI [1.04, 5.74]). Gen Z grew up with smartphones and overwhelmingly lock them, Gen X and older were adults when they emerged and are less likely to lock their phone. For Millenials on the boundary, an individual's level of familiarity with technology is a relevant predictor of their behavior. Those with little tech familiarity lock their phones less, while tech experts lock their phones more. For the proficient (but not expert) user in the middle, age is a relevant predictor of behavior.

\input{table_rel_age_lock}

Finally, there is a significant difference in locking behavior between operating systems, with 81\% of Android users (N = 253) and 92.2\% of iPhone users (N = 206) reporting that their phone locks ($\chi^2(8,460)=2.48, p=.013$). This means that, holding other factors constant, iOS users are 134\% more likely to lock their phones than Android users (95\% CI [1.22, 4.72]). The differences between the two operating systems cannot be explained by underlying demographic factors, and thus we should likely consider this correlation a reflection of a genuine phenomenon. This could be because of defaults and nudges, with iOS \textit{causing} more individuals to lock their phones, and/or it could be a result of Apple's marketing---\textit{"Privacy. That's iPhone."}---causing privacy conscious individuals to opt for iOS devices. 

\subsection*{Limitations of Authentication Schemes}

Many factors influence which authentication scheme(s) a person uses on their smartphone, most notably the options available on the device and the operating system (which we have provided a brief overview of below.)
For this reason, we have not performed extensive demographic analysis on the types of authentication in use.

iOS devices which lock require a PIN,
though users may replace it with an alphanumeric password.
Beginning with the iPhone 5S in 2013, Apple introduced TouchID, or the ability to unlock one's phone using a fingerprint reader~\cite{faceid}. 
With the iPhone X in 2017, Apple introduced FaceID, or the ability to unlock one's phone using facial recognition~\cite{touchid}. 
iPhones with TouchID or FaceID enabled are still required to have a PIN as a backup~\cite{still_need_pin}. 
Because Android runs on devices made by myriad manufacturers, it is less clear when certain authentication schemes became available. 
%Android devices default to a 4-digit PIN, though a longer PIN can be used as can an alphanumeric password. 
%Android also allows users to design a swipe pattern, drawing connections on a grid of 9 dots. 
%A swipe pattern may be used in place of a PIN. 
%Android devices may have fingerprint sensors, facial recognition, or iris readers, or some combination thereof; flagship Samsung devices have all three, while inexpensive devices marketed to the developing world offer no biometric authentication.

Devices with the option for biometric authentication still require a form of Something You Know (SYK) authentication be enabled as a backup because biomteric authentication can fail~\cite{bioproblems}. 
The SYK is either a PIN, password, or swipe pattern. 
It is possible to set up a device to recognize more than one biometric, e.g., more than one face or more than one finger, but it is critical to understand that the device does not have the capacity to distinguish between faces or fingers, and as such, any face or finger that has been given access on the device will have as much authority as any other~\cite{2fingers}.

%Both Android and iOS rely on a singular account with the OS' manufacturer---Google and Apple respectively---to enable and facilitate most features; for example, one cannot download apps without logging into one of these accounts. 
%If one does not change the default settings, (and most people don't \cite{}), these accounts permeate devices, backing up files, photos and messages to iCloud or Google Drive, monitoring the device's location and making that information searchable via Find My iPhone or Google Maps, and even storing a complete image of the device.
%so that if the user purchases a new phone, there is no need to set anything up. 
%While these features are convenient, they place a tremendous amount of value, and therefore vulnerability, with these keystone accounts.

For both iOS and Android, the SYK authentication is authoritative with respect to the device and operating system. 
The SYK functions as the cryptographic key for the hard drive, and on Android phones can be needed to launch the operating system. 
%after a restart or reset, and sensitive setting changes cannot be approved using biometric authentication. 
If authentication is required, it will be the device SYK, or the password for the underlying iCloud or Google account. 
Conversely, biometrics can be used as go-between authentication for sensitive apps such as banking apps, purchases through ApplePay or GooglePay, and approving app downloads~\footnote{By default, specific authentication is only required for non-free apps, but this setting can be changed.}, while SYKs cannot.

Of respondents whose phones lock (N = 396), 68.2\% had some form of biometric authentication enabled, including 84.2\% of iOS users (N = 184), and 56.4\% of Android users (N = 204). This difference is significant ($\chi^2(1, 396)=28.61, p<.0001$).
%All but five Android users with facial recognition enabled also had fingerprint sensing enabled, indicating that when multiple authentication options are made available to users, they take advantage of them. 
%No iOS device allows both fingerprint and facial recognition, and biometric authentication users in our sample (N = 155) were split about evenly between FaceID (53\%) and TouchID (47\%). 
%See Table \ref{tab:auth_os} for details.
Of respondents whose phones lock, 368 provided information about the form of SYK authentication in use.
PINs were the most common form of SYK used, accounting for 78.3\% of users, swipe patterns were also common, with 23.9\% of users having one, and 7.9\% of users reported using both. 
Further, 10.6\% of users reported using an alpha-numeric password.

%\input{table_auth_os}

%Beginning with iOS 9 in 2015 Apple has required a 6-digit PIN, replacing the previous 4-digit default.
%\footnote{It is understood that this change was in response to the US government's request for a backdoor, and subsequent use of cracking software after Apple's refusal to create one. The 6-digit PIN does not negate crackability, but increases the time needed to do so by 100x, making it intractable at scale. \cite{6digit_PIN}}

\section*{Access and Demographic Factors}~\label{app:access}
We used a logistic regression model to identify demographic factors correlated with sharing access to a user's phone, both implicitly and explicitly. The results of the implicit access model are in Table~\ref{tab:implicit_sharing_logreg}, and the results of the explicit access model are in Table~\ref{tab:explicit_sharing_logreg}.

\input{table_implicit_sharing_logreg}

\input{table_explicit_sharing_logreg}

Men are less likely to share access to their device than women, as they lock their phones significantly more often and share access less often. Including explicit and implicit access, 70.8\% of female respondents (N = 216) indicate that their partner has access to their device, as do 57\% of male respondents (N = 149). For implicit access, men are significantly less likely to share access ($\chi^2(6,402)=-2.41, p=.016$) by 59.6\%  (95\% CI [0.19, 0.82]). When considering only explicit access, there is a smaller difference between men (58.3\%) and women (69.8\%) ($\chi^2(6,402)=-1.52, p=.129$). 
These findings could suggest that there are true gendered differences with respect to preference for partner access, but that these preferences are mediated by desires for access parity as demonstrated in \ref{subsub:lock_demo}.

There is a significant difference in levels of explicit partner access between those in ``serious'' and ``less serious'' relationships. Those in ``serious'' relationships are not significantly less likely to lock their phones ($\chi^2(6, 402)=1.54, p=.123$), while there is a significant difference in the level of \textit{explicit} partner access between the two groups ($\chi^2(6, 402)=3.10, p=.0019$). One potential explanation is that those in ``serious'' relationships likely cohabitate with their partner, whereas people who are single or in ``less serious'' relationships may live with roommates or housemates.

Despite a significant increase in locking rates for Gen Z compared to other age groups, the rates at which those who lock their phones \textit{choose} to give their partners access are significantly higher ($\chi^2(6, 402)=3.54, p=.0004$): 68\% of Gen Z's partners had access, 93\% of which was given explicitly. This implies that Gen Z are not locking their phones to keep their partners out, but rather following better security practices, perhaps due to growing up surrounded by technology.

%% file: table_locking_logreg.tex
\begin{table*}[h]
\centering
\begin{tabular}{l|rrrrrrr}
                        & Coef   & SE    & p              &  OR   & Lower  & Upper \\ \hline
Gender - Male           &  0.849  & 0.362 & \textbf{0.019} & 2.34  & 1.17   & 4.89 \\
Generation - Gen Z      &  0.870  & 0.431 & \textbf{0.044} & 2.39  & 1.04   & 5.74 \\
Generation - Millenial  & -0.187 & 0.366 & 0.609          & 0.829 & 0.400  & 1.69 \\
Is LGBTQ                & 0.564  & 0.636 & 0.375          & 1.76  & 0.580  & 7.65 \\
iOS User                & 0.8508   & 0.343 & \textbf{0.013} & 2.34  & 1.22   & 4.72 \\
Technical Knowledge     & -0.2285  & 0.168 & 0.175          & 0.796 & 0.564  & 1.09
\end{tabular}

\caption{Logistic Regression Model Predicting if a Participant Locks Their Phone. Significant p-values are in \textbf{bold}}~\label{tab:lock_logreg}
\end{table*}                                      

%% file: table_rel_age_lock.tex
\begin{table*}[]
\centering
    \begin{threeparttable}
\begin{tabular}{l|ll|ll|ll|ll}
                                                       & \multicolumn{2}{l|}{Gen Z} & \multicolumn{2}{l|}{Millenial} & \multicolumn{2}{l|}{Gen X} & \multicolumn{2}{l}{Boomers} \\
                                                       & N             & \% Locks          & N         & \% Locks       & N         & \% Locks       & N        & \% Locks      \\ \hline
All                                                    & 171           & .936              & 154       & .851           & 103       & .816           & 32       & .656          \\ \hline
Single                                                 & 37            & .892              & 15        & .933           & 5         & 1              & 1        & 0             \\
\begin{tabular}[c]{@{}l@{}}Relationship\end{tabular}   & 134           & .948              & 139       & .842           & 98        & .806           & 31       & .677          \\ \hline
Serious                                                & 26            & .808              & 81        & .827           & 75        & .8             & 21       & .762          \\
\begin{tabular}[c]{@{}l@{}}Less Serious\end{tabular}   & 108           & .981              & 58        & .862           & 23        & .826           & 10       & .5      
\end{tabular}

%\begin{tablenotes}
%\footnotesize
%\item[***]Significant difference in locking behavior across age groups, $\chi^2(3, 460) = 21.08, p = .0001$
%\item[**]Significant difference in Gen Z between "serious" and "less serious" relationships, $\chi^2(1, 171)=10.22, p=.001$
%\end{tablenotes}
\end{threeparttable}

\caption{Rate of phone locking by age and relationship status}
~\label{tab:rel_age_lock}
\end{table*}

%% file: table_implicit_sharing_logreg.tex
\begin{table*}[h]
\centering
\begin{tabular}{l|rrrrrrr}
                        & Coef   & SE    & p               &  OR    & Lower  & Upper   \\ \hline
Gender - Male           & -0.905 & 0.376 & \textbf{0.0161} &  0.404 & 0.187  &   0.824 \\
Generation - Gen Z      & -0.934 & 0.541 & 0.0843          &  0.393 & 0.127  &   1.09  \\
Generation - Millenial  &  0.223 & 0.374 & 0.552           &  1.25  & 0.602  &   2.63  \\
Serious Relationship    &  0.575 & 0.373 & 0.123           &  1.78  & 0.869  &   3.78  \\
Is LGBTQ                & -0.459 & 0.646 & 0.477           &  0.632 & 0.143  &   1.97  \\
iOS User                & -0.985 & 0.377 & \textbf{0.0089} &  0.374 & 0.171  &   0.759
\end{tabular}

\caption{Logistic Regression Model Predicting if a Participant \textit{Implicitly} Shares Phone Access. Significant p-values are in \textbf{bold}}
~\label{tab:implicit_sharing_logreg}
\end{table*}

%% file: table_explicit_sharing_logreg.tex
\begin{table*}[h]
\centering
\begin{tabular}{l|rrrrrrr}
                       & Coef   & SE    & p               &  OR   & Lower & Upper  \\ \hline
Gender - Male          & -0.359 & 0.236 & 0.129           & 0.699 & 0.439 & 1.11 \\
Generation - Gen Z     &  1.219 & 0.344 & \textbf{0.0004} & 3.38  & 1.74  & 6.73 \\
Generation - Millenial &  0.384 & 0.295 & 0.193           & 1.47  & 0.826 & 2.63 \\
Serious Relationship   &  0.821 & 0.265 & \textbf{0.0019} & 2.27  & 1.36  & 3.86 \\
Is LGBTQ               &  0.201 & 0.344 & 0.559           & 1.22  & 0.626 & 2.43 \\
iOS User               &  0.096 & 0.228 & 0.674           & 1.10  & 0.703 & 1.72 
\end{tabular}

\caption{Logistic Regression Model Predicting if a Participant \textit{Explicitly} Shares Phone Access. Significant p-values are in \textbf{bold}}
~\label{tab:explicit_sharing_logreg}
\end{table*}